\documentclass[a4paper,11pt]{article}

\usepackage{jheppub} 
\usepackage[T1]{fontenc} 
\usepackage{slashed}
\usepackage[utf8x]{inputenc}
\usepackage{color}
\usepackage[normalem]{ulem}
\usepackage{bm}

\allowdisplaybreaks

\title{CP asymmetry in heavy Majorana neutrino decays at finite temperature: the hierarchical case}

\author[a]{S. Biondini,}
\author[a,b]{N. Brambilla,}
\author[a]{and A. Vairo}

\affiliation[a]{Physik-Department, Technische Universit\"{a}t M\"{u}nchen,\\ James-Franck-Str. 1, 85748 Garching, Germany}
\affiliation[b]{Institute for Advanced Study, Technische Universit\"{a}t M\"{u}nchen,\\ Lichtenbergstrasse 2 a, 85748 Garching, Germany}

\emailAdd{simone.biondini@ph.tum.de}
\emailAdd{nora.brambilla@ph.tum.de}
\emailAdd{antonio.vairo@ph.tum.de}

\abstract{We consider the simplest realization of leptogenesis with one heavy Majorana neutrino species much lighter than the other ones.
In this scenario, when the temperature of the early universe is smaller than the lightest Majorana neutrino mass, 
we compute at first order in the Standard Model couplings and, for each coupling, at leading order in the termperature 
the CP asymmetry in the decays of the lightest neutrino into leptons and anti-leptons. 
We perform the calculation using a hierarchy of two effective field theories organized as expansions in the inverse of the heavy-neutrino masses. 
In the ultimate effective field theory, leading thermal corrections proportional to the Higgs self coupling 
and the gauge couplings are encoded in one single operator of dimension five, whereas corrections proportional to the top Yukawa coupling 
are encoded in four operators of dimension seven, which we compute.}


\begin{document} 
\maketitle
\flushbottom

\section{Introduction}
The explanation of the observed baryon asymmetry in the universe poses
an interesting and challenging task to cosmology and particle physics. 
Since any ab-initio imbalance between particles and anti-particles in the very early universe 
has been likely washed-out after the inflationary epoch, a dynamical generation of the
present baryon asymmetry, or baryogenesis, appears favoured.

Baryogenesis requires typically out-of-equilibrium decays of heavy particles. 
A first realization was proposed within the framework 
of the Grand Unified Theories (GUTs)~\cite{Yoshimura:1978ex,Toussaint:1978br,Weinberg:1979bt,Dimopoulos:1978kv}. 
The heavy gauge bosons predicted in the GUTs, with masses of the order of
10$^{15}$-10$^{16}$~GeV, are the source of the baryon asymmetry once
baryon number, C and CP violating processes are introduced in the model~\cite{Sakharov:1967dj}. 
The decays of these heavy states produce different amounts of particles and anti-particles providing the desired imbalance. 
Two major issues affect this scenario. 
First, the final asymmetry depends on too many free parameters limiting the predictive power. 
Second, the reheating temperature after the inflationary epoch cannot be higher than 10$^{15}$~GeV 
as accounted for by the Cosmic Microwave Background analysis~\cite{Giudice:2000ex}, 
which could affect the thermal production of the heavy particles predicted by the GUTs 
undermining the very basis of such a scenario~\cite{Blanchet:2012bk}.

On the other hand, baryogenesis via leptogenesis~\cite{Fukugita:1986hr} is an attractive class of models that avoids some of the issues related to GUT baryogenesis. 
The original and minimal version of leptogenesis requires heavy right-handed neutrinos in addition to the Standard Model (SM) particles.
Right-handed neutrinos may be embedded into Majorana fields.
Because of the CP-violating phases of their Yukawa couplings with Higgs bosons and leptons, 
they decay into different amounts of leptons and anti-leptons.
Sphaleron transitions convert eventually the lepton asymmetry into a baryon asymmetry~\cite{Kuzmin:1985mm}. 
Moreover, heavy neutrinos may participate in the type I seesaw mechanism~\cite{Minkowski:1977sc,GellMann:1980vs,Mohapatra:1979ia}, 
providing a natural explanation of the small masses of the three SM neutrinos. 
Indeed the discovery of the neutrino oscillations and mixing has shown that neutrinos do have masses~\cite{Fukuda:1998mi} 
and some mechanism that generates such masses is necessary. 
Also, the solution of the Boltzmann equations provides hints to the highest temperature needed for a successful leptogenesis. 
This is found to be up to ten times lower than the reheating temperature after inflation, depending on the values of the
Yukawa couplings among heavy neutrinos and SM Higgs bosons and leptons~\cite{Buchmuller:2004nz}.

We will not discuss further neither the theoretical foundation and mechanism of leptogenesis 
nor the phenomenology of right-handed/Majorana neutrinos, which are widely addressed in exhaustive reviews, 
e.g., in~\cite{Fong:2013wr,Buchmuller:2005eh} and~\cite{Drewes:2013gca}. 
Here we will focus on one particular aspect. 
Since the heavy-neutrino dynamics occurs in a thermalized medium made of SM particles, 
namely the universe in its early stages, we will study the impact of thermal effects
on the CP asymmetry originated in the neutrino leptonic decays.
When the temperature is smaller than the neutrino masses one may exploit this hierarchy 
to construct suitable effective field theories (EFTs) and compute observables in a systematic 
expansion in the inverse of the neutrino masses~\cite{Biondini:2013xua}.
In this framework, we have recently derived the CP asymmetry at finite temperature 
for the case of two heavy Majorana neutrinos with nearly degenerate masses~\cite{Biondini:2015gyw}.
In the present work, we compute the leading thermal corrections to the CP asymmetry 
for the case of a hierarchically ordered mass spectrum of Majorana neutrinos.

Some finite temperature studies of the CP asymmetry can be found in~\cite{Covi:1997dr,Giudice:2003jh}. 
Several investigations of the lepton-number asymmetry have been carried out either within the
Boltzmann rate equations and their quantum version known as Kadanoff--Baym equations~\cite{Garny:2010nj,Anisimov:2010dk,Kiessig:2011fw}. 
Thermal effects are typically accounted for by including thermal masses and
thermal distributions for the Higgs bosons and leptons appearing as decay products of heavy Majorana neutrinos. 
In the present work, we provide a systematic derivation of the thermal corrections 
to the CP asymmetry in terms of an expansion in the SM couplings and in $T/M_I$,
where $M_I$ are the masses of the heavy Majorana neutrinos and $T$ is the temperature.

More precisely we consider the simplest realization of leptogenesis often called \emph{vanilla leptogenesis} in the literature.  
In this scenario one assumes one Majorana neutrino, with mass $M_1$, much lighter than the other heavy neutrinos. 
Under this assumption, the final CP asymmetry is produced by the lightest neutrino decays. 
Moreover, we assume that different lepton (anti-lepton) flavours are resolved by the thermal bath during leptogenesis. 
This regime is called flavoured in contrast to the unflavoured regime 
that describes the situation when the different flavours are not resolved by the thermal bath.
The flavoured regime applies to a larger range of temperatures than the unflavoured one.
For instance, the three lepton flavours are resolved by charged Yukawa coupling interactions already at temperatures 
of the order of $10^9$~GeV~\cite{Nardi:2005hs,Nardi:2006fx}, whereas the unflavoured regime is found to be an appropriate choice 
only at very high temperatures, in particular $T \gtrsim 10^{12}$~GeV. 
In the flavoured case it makes sense to define a CP asymmetry for each lepton flavour; 
the CP asymmetry generated by the lightest Majorana neutrino decaying into leptons and anti-leptons of flavour $f$ reads
\begin{equation}
\epsilon_{f}=  \frac{  \Gamma(\nu_{R,1} \to \ell_{f} + X)-\Gamma(\nu_{R,1} \to \bar{\ell}_{f}+ X )  }
{\sum_f \, \Gamma(\nu_{R,1} \to \ell_{f} + X )+\Gamma(\nu_{R,1} \to \bar{\ell}_{f}+ X)} \,.
\label{CPdef1}
\end{equation}
In \eqref{CPdef1} $\nu_{R,1}$ stands for the lightest right-handed/Majorana neutrino, 
$\ell_f$ is a SM lepton with flavour $f$ and $X$ represents any other SM particle not carrying a lepton number.  
When summing over the flavours also in the numerator of \eqref{CPdef1}, 
we recover the CP asymmetry in the unflavoured vanilla leptogenesis scenario. 
In this scenario, experiments looking at neutrino oscillations and mixing parameters 
can put constraints on some of the leptogenesis parameters. 
An example is the Davidson--Ibarra bound that provides a lower bound on the lightest
heavy-neutrino mass~\cite{Davidson:2002qv, Buchmuller:2002rq}, $M_1 \gtrsim 10^{9}$ GeV. 
It is obtained combining the observed baryon asymmetry and the light neutrino masses. 
This bound sets the energy scale of leptogenesis, at least in its simplest realization, 
together with the typical temperatures needed for the heavy-neutrino thermal production.
In the flavoured regime, the lower bound on the lightest Majorana neutrino mass can be relaxed down to $M_1 \gtrsim 10^{6}$~GeV, 
due to modifications of the heavy-neutrino dynamics induced by different flavour effects~\cite{Racker:2012vw}.

A crucial transition for the generation of the lepton asymmetry happens when the temperature of the thermal plasma, $T$, 
equals the mass of the lightest Majorana neutrino: $T \sim M_1$. 
In fact, while for $T > M_1$, the originated CP asymmetry can be efficiently erased if the so-called strong wash-out is assumed, 
which seems to be the favoured scenario according to the present values of solar and atmospheric neutrino oscillation data,
this is no more the case for $T < M_1$. 
Hence, the final asymmetry turns out to be independent of the initial abundance of the lightest Majorana neutrinos
and is effectively generated at temperatures smaller than the lightest neutrino mass~\cite{Buchmuller:2004nz,Blanchet:2012bk}. 
For $T < M_1$, the lightest neutrino is out-of-equilibrium with the thermal bath, 
a necessary condition for the matter-antimatter asymmetry. Moreover, its dynamics is non-relativistic.

We consider three species of heavy neutrinos, though in general the model may account for a generic number of species.\footnote{
At least two heavy-neutrino species are necessary to have non-vanishing CP asymmetries.} 
We call $M_1$ the mass of the lightest right-handed neutrino and $M_i$, $i =2,3$, the masses of the heavier neutrinos.
Moreover, we assume the temperature, $T$, of the thermal plasma in the early universe 
to be much smaller than the mass of the lightest neutrino and larger than the electroweak scale, $M_W$. 
This means that we assume the following hierarchy of energy scales\footnote{
Thermal modes are associated to the Matsubara frequencies of the plasma, 
hence the relevant thermodynamical scale is proportional to $\pi T$. 
Through the paper we will assume $T$ and $\pi T$ to be parametrically equivalent scales.
We will restore the scale $\pi T$ in figure~\ref{fig_plotCPcontributions}.
}  
\begin{equation} 
M_{i} \gg M_{1} \gg T \gg M_W \,, \quad \hbox{for}\;\; i=2,3\,.
\label{Hiera}
\end{equation} 
The last inequality, setting the temperature above the electroweak scale, 
ensures that the SM sector is described by an unbroken SU(2)$_L \times$U(1)$_Y$ gauge symmetry, 
which implies that all SM particles are massless. 

\begin{figure}[ht]
\centering
\includegraphics[scale=0.65]{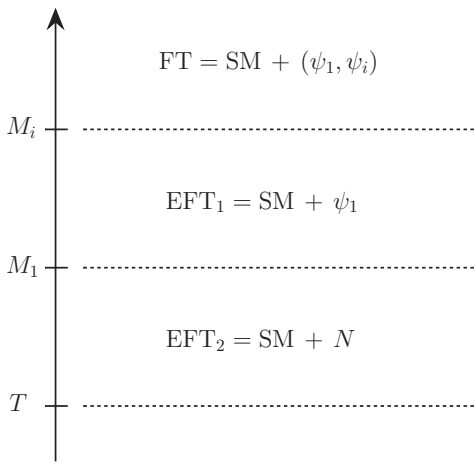}
\caption{The relevant hierarchy of energy scales is shown together with the corresponding hierarchy of EFTs.
FT stands for the fundamental theory \eqref{Lag1},
whose degrees of freedom are the SM particles and all three species of heavy Majorana neutrinos ($\psi_1$, $\psi_i$).
By integrating out the scales $M_i$ and $M_1$ one obtains sequentially the EFTs: EFT$_1$ \eqref{Lag2} and EFT$_2$ \eqref{Lag3}. 
In the former the degrees of freedom are the SM particles and the lightest Majorana neutrino ($\psi_1$), 
whereas in the latter only SM particles and non-relativistic modes of the lightest neutrino ($N$) are dynamical.}
\label{fig:EFTs}
\end{figure}

We exploit the hierarchy of energy scales \eqref{Hiera} by constructing a hierarchy of two EFTs.
In a first EFT, we integrate out modes with energy and momentum of the order of the heavier neutrino masses, $M_i$. 
The degrees of freedom of the EFT are the SM particles and the lightest Majorana neutrino. 
The EFT contains effective vertices between SM leptons and Higgs bosons~\cite{Buchmuller:2000nd}. 
We call it EFT$_1$ throughout the paper.
In a second EFT, we integrate out modes with energy and momentum of the order of the lightest neutrino mass,~$M_1$.
The degrees of freedom are the SM particles and the non-relativistic modes of the lightest Majorana neutrino,  
which appears as an initial state in the observable that we compute.
We call this second EFT, EFT$_2$.  The hierarchy of EFTs is shown in figure~\ref{fig:EFTs}.

In the paper, we compute the leading thermal corrections to the CP asymmetry in the leptonic decays of the lightest Majorana neutrino 
at first order in the SM couplings. The most suitable EFT for performing this computation is EFT$_{2}$. 
The calculation can be done using the techniques developed in~\cite{Biondini:2013xua,Biondini:2015gyw}. 
Moreover, some of the results may be checked against intermediate expressions obtained in~\cite{Biondini:2015gyw}.
At first order in the Higgs self-coupling and in the gauge couplings,  
the leading thermal correction to the CP asymmetry is of relative order $(T/M_1)^2$ 
and encoded in one dimension-five operator of the EFT$_{2}$.
At first order in the top-quark Yukawa coupling,  
the leading thermal correction to the CP asymmetry is of relative order $(T/M_1)^4$ 
and encoded in four dimension-seven operators of the EFT$_{2}$. 
The dimension-five and -seven operators were identified in~\cite{Biondini:2013xua}, 
but here we need to compute the contributions to their Wilson coefficients that are relevant for the CP asymmetry. 
This computation is new.  Since $T \ll M_1$ the matching of the Wilson coefficients can be done setting the temperature to zero. 
This amounts at evaluating two-loop cut diagrams in vacuum matching the dimension-five and -seven operators. 
Once the Wilson coefficients are known, thermal corrections are encoded in the thermal expectation values of the corresponding operators. 
Their computation requires that of a simple tadpole diagram.
The final expression of the CP asymmetry follows from the definition \eqref{CPdef1}.

Thermal corrections to the CP asymmetry and to the heavy-neutrino production rate enter 
the rate equations for the heavy-neutrino and lepton-asymmetry number densities. 
Thermal corrections to the right-handed neutrino production rate have been derived in~\cite{Laine:2013lka} 
for the relativistic and ultra-relativistic regimes, 
whereas the non-relativistic case has been addressed in~\cite{Salvio:2011sf,Laine:2011pq} and \cite{Biondini:2013xua}. 
In order to connect those results with leptogenesis, 
Boltzmann equations in the non-relativistic regime have been derived in~\cite{Bodeker:2013qaa}.
The thermally corrected production rate has been used to solve the rate equations for the out-of-equilibrium dynamics.
Studies in this direction may be further improved by using the thermally corrected expression for the CP asymmetry that we compute here.

The paper is organized as follows. 
In section~\ref{sec1} we summarize the results for the CP asymmetry at zero temperature. 
In section~\ref{sec_eft1} we derive the EFT$_{1}$ (details can be found in appendix~\ref{AppA}).
The most original results of the paper are in sections~\ref{sec2} and~\ref{sec3}. 
In section~\ref{sec2} we build the EFT$_{2}$ and compute the relevant Wilson coefficients (details of the matching are in appendix~\ref{AppB}).
In section~\ref{sec3} we derive the thermal corrections to the CP asymmetry. 
Conclusions and discussions are collected in section~\ref{sec4}.

\section{CP asymmetry at zero temperature}
\label{sec1}
We consider an extension of the SM that includes three heavy Majorana neutrinos coupled to the SM Higgs boson and lepton doublets. 
The Lagrangian of our fundamental theory reads~\cite{Fukugita:1986hr}
\begin{equation}
\mathcal{L}=\mathcal{L}_{\hbox{\tiny SM}} 
+ \frac{1}{2} \,\bar{\psi}_{I} \,i \slashed{\partial}  \, \psi_{I}  - \frac{M_I}{2} \,\bar{\psi}_{I}\psi_{ I} 
- F_{fI}\,\bar{L}_{f} \tilde{\phi} P_{R}\psi_{I}  - F^{*}_{fI}\,\bar{\psi}_{I}P_{L} \tilde{\phi}^{\dagger}  L_{f} \, ,
\label{Lag1}
\end{equation} 
where $\mathcal{L}_{\hbox{\tiny SM}} $ is the SM Lagrangian \eqref{SMlag}, 
$\psi_I = \nu_{R,I} + \nu_{R,I}^c$ stands for the $I$-th Majorana field embedding the right-handed neutrino field $\nu_{R,I}$, 
with mass $M_I$ and $I=1,i\; (i=2,3)$ is the mass eigenstate index.  
The fields $L_{f}$ are lepton doublets with flavour $f$, $\tilde{\phi}=i \sigma^{2} \, \phi^*$, 
where $\phi$ is the Higgs doublet, and $F_{fI}$ is a (complex) Yukawa coupling.
The left-handed and right-handed projectors are $P_L = (1 - \gamma^5)/2$ and $P_R = (1 + \gamma^5)/2$ respectively.    
We consider the case of one heavy Majorana neutrino species much lighter than the other ones: $M_1 \ll M_i$.

\begin{figure}[ht]
\centering
\includegraphics[scale=0.55]{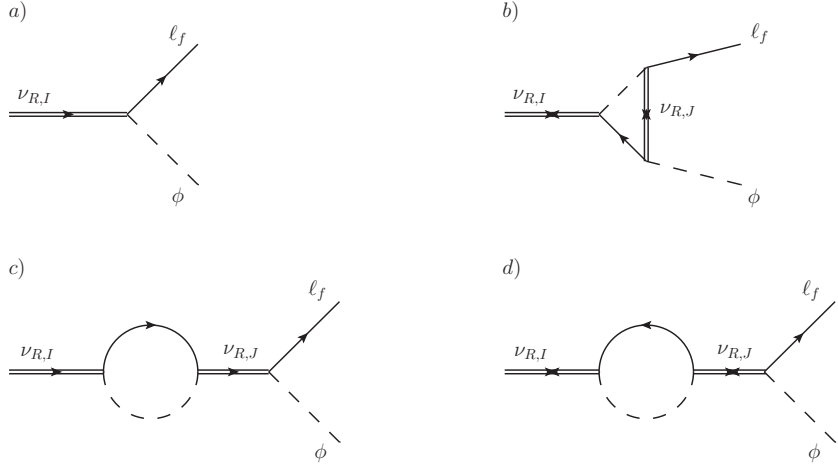}
\caption{CP asymmetry is originated from the interference between tree-level and one-loop vertex and self-energy (or wave-function) diagrams. 
Solid double lines stand for heavy right-handed neutrinos, solid lines for SM lepton doublets and dashed lines for Higgs bosons. 
The neutrino propagator with forward arrow corresponds to $\langle 0| T(\psi \bar{\psi}) |0\rangle$, 
whereas the neutrino propagators with forward-backward
arrows correspond to $\langle 0| T(\psi \psi) |0\rangle$ or $\langle 0| T(\bar{\psi} \bar{\psi})|0 \rangle$.}
\label{fig:fig1} 
\end{figure}

The CP asymmetry at zero temperature can be calculated at leading order from the interference between the tree-level and one-loop diagrams 
shown in figure~\ref{fig:fig1}. 
Diagram $b)$ is referred to as the vertex diagram, whereas diagrams $c)$ and $d)$ are often called self-energy diagrams, 
diagram $c)$ being relevant only for the flavoured CP asymmetry.
Their contribution to the CP asymmetry depends on the heavy-neutrino mass spectrum. 
It is known that in the case of a hierarchical neutrino mass spectrum
the two contributions are of the same order and, in particular, the one
originated by the self-energy diagram is twice as big as the vertex one~\cite{Liu:1993tg,Covi:1996wh}.  

\begin{figure}[ht] \centering
\includegraphics[scale=0.55]{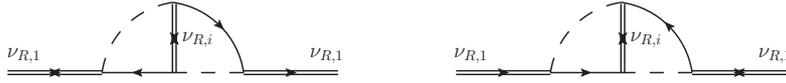}
\caption{Self-energy diagrams for the lightest Majorana neutrino, $\nu_{R,1}$, corresponding to the mass eigenstate with mass $M_1$. 
The imaginary parts of the diagrams provide the interference between the tree-level and the one-loop vertex diagram in figure~\ref{fig:fig1}.}
\label{fig:self1} 
\end{figure}

\begin{figure}[ht]  \centering
\includegraphics[scale=0.52]{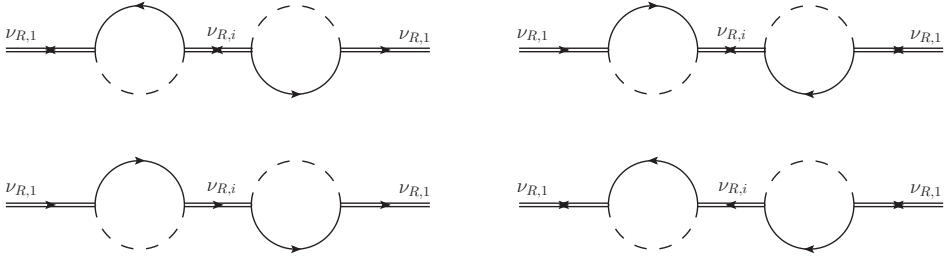}
\caption{Self-energy diagrams for the lightest Majorana neutrino. 
The imaginary parts of the diagrams provide the interference between the tree-level and the one-loop self-energy diagrams in figure~\ref{fig:fig1}.}
\label{fig:self2} 
\end{figure}

The interference between the tree-level and one-loop diagrams in figure~\ref{fig:fig1} may be computed 
from the imaginary part of the heavy-neutrino self-energy at fourth-order in the Yukawa couplings. 
We have presented in detail how this works for the vertex topology in the nearly degenerate case in~\cite{Biondini:2015gyw},
including also the treatment of flavour effects. 
In the hierarchical case we may use the same arguments to write the CP asymmetry \eqref{CPdef1} 
for the decays into lepton species of flavour $f$ due to the vertex diagram, $\epsilon_{f,{\rm{direct}}}$, 
and due to the self-energy diagram, $\epsilon_{f,{\rm{indirect}}}$, in the general form 
\begin{eqnarray}
\epsilon_{f}= \epsilon_{f,{\rm{direct}}} + \epsilon_{f,{\rm{indirect}}} = 
&&- 2 \sum_{I} {\rm{Im}}(B_{\hbox{\tiny direct}}+ B_{\hbox{\tiny indirect}})  \frac{{\rm{Im}} \left[ (F_{1}^{*}F_{I})(F^*_{f1}F_{fI}) \right] }{|F_{1}|^2}
\nonumber
\\
&&- 2 \sum_{I} {\rm{Im}}(\tilde{B}_{\hbox{\tiny indirect}})  \frac{{\rm{Im}} \left[ (F_{1}F_{I}^{*})(F^*_{f1}F_{fI}) \right] }{|F_{1}|^2} ,
\label{CPdef2}
\end{eqnarray}
where  $(F_{1}^{*}F_{I}) \equiv \sum_f F^*_{f1}F_{fI}$. 
The functions $B_{\hbox{\tiny direct}}$, $B_{\hbox{\tiny indirect}}$ and $\tilde{B}_{\hbox{\tiny indirect}}$ can be calculated
by cutting the two-loop diagrams shown in figure~\ref{fig:self1} and~\ref{fig:self2}, first and second raw, respectively. 
These diagrams contribute to the propagator of the lightest Majorana neutrino
\begin{equation} 
-i \left. \int d^{4}x \, e^{ip\cdot x} \, \langle
\Omega | T \left( \psi_{1}^{\mu}(x) \bar{\psi}_{1}^{\nu}(0) \right) | \Omega \rangle \right|_{p^\alpha =(M_1 + i\epsilon,\vec{0}\,)} \, ,
\label{matrixFund}
\end{equation} 
where $| \Omega \rangle$ stands for the ground state of the fundamental theory. 
The term in the second line in \eqref{CPdef2}, which originates from the two diagrams in the lower row 
of figure~\ref{fig:self2}, vanishes in the unflavoured regime because $\sum_f {\rm{Im}} [ (F_{1}F_{I}^{*})(F^*_{f1}F_{fI}) ]=0$.

At zero temperature the CP asymmetry induced by the diagrams in figure~\ref{fig:self1} is given by~\cite{Covi:1996wh,Fong:2013wr}
\begin{eqnarray}
\epsilon_{f,{\rm{direct}}}^{T=0} &=& \frac{M_i}{M_1} \left[ 1-\left( 1+\frac{M^2_i}{M^2_1} \right)  
\ln \left( 1+ \frac{M_1^2}{M_i^2} \right) \right] \frac{ {\rm{Im}}\left[ (F_{1}^{*}F_{i})(F^*_{f1}F_{fi})  \right] }{8 \pi|F_1|^2} 
\nonumber \\
&\underset{M_1 \ll M_i}{=}& -\frac{1}{16 \pi } \frac{M_1}{M_i}\frac{ {\rm{Im}}\left[(F_{1}^{*}F_{i})(F^*_{f1}F_{fi})  \right] }{|F_1|^2} 
+\mathcal{O}\left( \frac{M_1}{M_i}\right)^3 \,.
\label{CPvert}
\end{eqnarray}
A sum over the intermediate heavy Majorana neutrino species, labeled by $i$ ($i =2,3$), is understood (this will be always the case in the following,  
if not specified differently); note, however, that we do not sum over the flavour, $f$, of the leptons. 

The CP asymmetry generated at zero temperature by the diagrams in figure~\ref{fig:self2} is~\cite{Covi:1996wh,Fong:2013wr}
\begin{eqnarray}
\epsilon_{f,{\rm{indirect}}}^{T=0}&=& \frac{M_1 M_i}{M_1^2-M_i^2} \frac{ {\rm{Im}}\left[ (F_{1}^{*}F_{i})(F^*_{f1}F_{fi})  \right] }{8 \pi|F_1|^2} 
+ \frac{M_1^2}{M_1^2-M_i^2} \frac{ {\rm{Im}}\left[ (F_{1}F_{i}^{*})(F^*_{f1}F_{fi})  \right] }{8 \pi|F_1|^2} 
\nonumber \\
&\underset{M_1 \ll M_i}{=}&-\frac{1}{8 \pi } \frac{M_1}{M_i}\frac{ {\rm{Im}}\left[(F_{1}^{*}F_{i})(F^*_{f1}F_{fi})  \right] }{|F_1|^2}  
-\frac{1}{8 \pi } \left(  \frac{M_1}{M_i}\right)^2\frac{ {\rm{Im}}\left[(F_{1}F_{i}^{*})(F^*_{f1}F_{fi})  \right] }{|F_1|^2} 
\nonumber \\
&&+ \,  \mathcal{O}\left( \frac{M_1}{M_i}\right)^3  .
\label{CPself}
\end{eqnarray}
In \eqref{CPself} the combination ${\rm{Im}}[(F_{1}^{*}F_{i})(F^*_{f1}F_{fi})]$  is originated by the upper-raw diagrams in figure~\ref{fig:self2}, 
whereas ${\rm{Im}}[(F_{1}F_{i}^{*})(F^*_{f1}F_{fi}) ]$ comes from the lower-raw diagrams. 
The latter combination, which contributes at order $(M_1/M_i)^2$, vanishes in the unflavoured regime. 
The assumption $M_1 \ll M_i$ selects implicitly a situation where the neutrino mass difference, $M_i-M_1$, 
is much larger than the heavy neutrino widths and mixing terms, preventing a resonant behaviour from happening.
We note that $\epsilon_{f,{\rm{indirect}}}^{T=0} = 2\epsilon_{f,{\rm{direct}}}^{T=0}$ at first order in~$M_1/M_i$.

\section{EFT$_1$}
\label{sec_eft1}
As our first task we derive the EFT$_1$, which is the EFT that follows from the fundamental theory \eqref{Lag1} 
by integrating out degrees of freedom with energies and momenta of order $M_i \gg M_1$ ($i =2,3$).
The EFT$_{1}$ will be our starting point for the construction of the EFT$_2$,  
where only degrees of freedom with energies and momenta smaller than $M_1$, the lightest Majorana neutrino mass, 
remain active. EFT$_2$ will be derived in section~\ref{sec2}.

Since we assume the temperature to be much smaller than the heavy-neutrino masses, see \eqref{Hiera}, 
we can set it to zero in the matching between the full theory \eqref{Lag1} and the EFT$_1$. 
Moreover, momenta and energies of external particles (in our case Higgs bosons and leptons) are taken much smaller than the masses $M_i$. 
The relevant operators to match are dimension-five and -six two-Higgs-two-lepton operators. 
Indeed, looking at the diagrams in the figures~\ref{fig:self1} and \ref{fig:self2}, we see 
that the intermediate interaction involving the heavy Majorana neutrinos with masses $M_i$ 
reduces to an effective two-Higgs-two-lepton vertex if we cannot resolve energies of the order of $M_i$ or higher.
At the accuracy that we compute the CP asymmetry, we do not need to match loop diagrams to the~EFT$_1$. 

\begin{figure}[ht]
\centering
\includegraphics[scale=0.58]{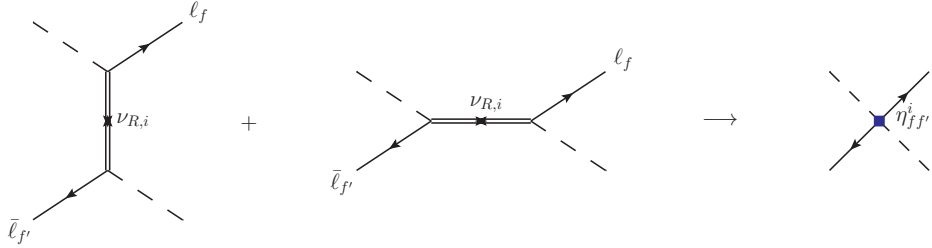}
\caption{Tree-level matching between the fundamental theory and a two-Higgs-two-lepton vertex of the EFT$_1$. 
The two diagrams in the left-hand side are the $t$-channel and $s$-channel interactions appearing 
in the diagrams of figures~\ref{fig:self1} and~\ref{fig:self2} (upper raw). 
In the right-hand side, the four-particle diagram stands for the effective two-Higgs-two-lepton interaction in the EFT$_1$.}
\label{fig:eft1} 
\end{figure}

In figure~\ref{fig:eft1} we illustrate the matching of the dimension-five two-Higgs-two-lepton operator in the EFT$_1$.
The left-hand side shows the lepton-number violating scattering $\bar{\ell} + \phi \to \ell + \phi$ mediated by heavy neutrinos of mass $M_i$ 
both in the $t$- and $s$-channels (the diagrams with the anti-lepton (lepton) outgoing (ingoing) are not shown, 
but contribute to the Hermitian conjugate operator). 

\begin{figure}[ht]
\centering
\includegraphics[scale=0.58]{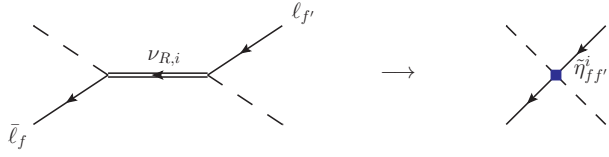}
\caption{Tree-level matching between the fundamental theory and a two-Higgs-two lepton vertex of the EFT$_1$. 
The diagram in the left-hand side is the $s$-channel interaction appearing in the diagrams of figure~\ref{fig:self2} (lower raw). 
In the right-hand side, the four-particle diagram stands for the effective two-Higgs-two-lepton interaction in the EFT$_1$.}
\label{fig:eft1_bis} 
\end{figure}

In figure~\ref{fig:eft1_bis} we illustrate the matching of the dimension-six two-Higgs-two-lepton operator in the EFT$_1$.
The left-hand side shows the lepton-number conserving scattering $\bar{\ell} + \phi \to \bar{\ell} + \phi$ 
mediated by heavy neutrinos of mass $M_i$ in the $s$-channel. 
The dimension-six operator in the EFT$_1$ in the right-hand side depends on the momentum of the anti-lepton-Higgs-boson pair.
A detailed account of the matching can be found in appendix~\ref{AppA}. 

\begin{figure}[ht]
\centering
\includegraphics[scale=0.6]{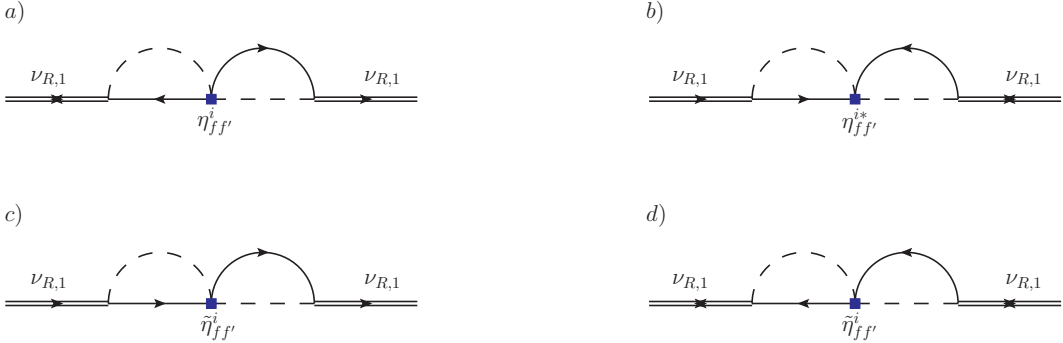}
\caption{Two-loop self-energy diagrams for the lightest neutrino, $\nu_{R,1}$, in the EFT$_{1}$.
The blue squared vertices correspond to the effective vertices of figure~\ref{fig:eft1} and~\ref{fig:eft1_bis}.
Diagram $a)$ reproduces the first diagram of figure~\ref{fig:self1} and the first diagram of figure~\ref{fig:self2}.
Diagram $b)$ reproduces the second diagram of figure~\ref{fig:self1} and the second diagram of figure~\ref{fig:self2}.
Diagram $c)$ reproduces the third diagram of figure~\ref{fig:self2} and diagram $d)$ the fourth diagram of figure~\ref{fig:self2}.}
\label{fig:eft1cp} 
\end{figure}

The difference between vertex and self-energy diagrams in the fundamental theory amounts to a
difference in the kinematical channel of the exchanged neutrinos of mass~$M_i$.
Specifically, an exchanged neutrino in the $t$-channel identifies a vertex diagram 
and an exchanged neutrino in the $s$-channel identifies a self-energy one.
For in the EFT$_1$ we cannot resolve the exchanged neutrinos, these two kinds of diagrams become indistinguishable. 
This is best shown in figure~\ref{fig:eft1}, where both type of diagrams contribute to the very same effective vertex.
As a consequence, at the level of the EFT$_1$ we cannot distinguish anymore 
between direct and indirect contributions to the CP asymmetry. 
In figure~\ref{fig:eft1cp} we reproduce in the EFT$_1$, up to order $(M_1/M_i)^2$, the diagrams in the fundamental theory 
shown in figure~\ref{fig:self1} and~\ref{fig:self2}. They will be computed in appendix~\ref{AppA1}, see figures~\ref{fig:appA_1} and~\ref{fig:appA_1_2}.

The EFT$_1$ Lagrangian including the dimension-five and  -six two-Higgs-two lepton operators matched 
in figure~\ref{fig:eft1} and~\ref{fig:eft1_bis} respectively reads 
\begin{eqnarray}
\mathcal{L}_{{\rm{EFT}_1}}=\mathcal{L}_{\hbox{\tiny SM}} 
&+& \frac{1}{2} \,\bar{\psi}_{1} \,i \slashed{\partial}  \, \psi_{1}  
- \frac{M_1}{2} \,\bar{\psi}_{1}\psi_{1}  - F_{f1}\,\bar{L}_{f} \tilde{\phi} P_{R}\psi_{1}  
- F^{*}_{f1}\,\bar{\psi}_{1}P_{L} \tilde{\phi}^{\dagger}  L_{f}
\nonumber \\
&+& \left( \frac{\eta_{ff'}^i}{M_i} \bar{L}_f \tilde{\phi} \, C P_R \,  \tilde{\phi}^T  \bar{L}^T_{f'} +  {\rm H.\,c.}\right) 
+ \frac{\tilde{\eta}^i_{ff'}}{M^2_i} \bar{L}_f  \tilde{\phi}  P_R \, i \slashed{\partial} (\tilde{\phi}^\dagger L_{f'})+  \dots \, ,
\label{Lag2}
\end{eqnarray}
where $C$ is the charge conjugation matrix, $\rm H.\,c.$ stands for Hermitian conjugate, $T$ for transpose 
and the dots for higher-order terms in the $1/M_i$ expansion. 
The coefficients $\eta_{ff'}^i$ and $\tilde{\eta}_{ff'}^i$ are the Wilson coefficients 
of the dimension-five (lepton-number violating) and dimension-six (lepton-number conserving) operators respectively. 
At leading order they read (from figures~\ref{fig:eft1} and~\ref{fig:eft1_bis} and appendix~\ref{AppA})
\begin{equation}
\eta^i_{ff'}=\frac{1}{2} F_{fi}F_{f'i} \, , \quad\qquad \tilde{\eta}^i_{ff'} =  F_{fi}F^*_{f'i}\, , 
\label{eq8}
\end{equation}
where, in this case, the index $i$ is not summed on the right-hand side of each Wilson coefficient. 
Note that the Lagrangian \eqref{Lag2} contains as degrees of freedom only the SM fields 
and the lightest Majorana neutrino field, $\psi_1$.

Within the EFT$_{1}$ one may reproduce the sum of the CP asymmetries \eqref{CPvert} and \eqref{CPself}, 
$\epsilon_{f}^{T=0} = \epsilon_{f,{\rm{direct}}}^{T=0} + \epsilon_{f,{\rm{indirect}}}^{T=0}$,  
order by order in $1/M_i$, see appendix~\ref{AppA1} and equation~\eqref{CPhiera}.  
This was first realized in~\cite{Buchmuller:2000nd}, where the EFT$_1$ Lagrangian and the CP asymmetry 
were computed up to order~$1/M_i$.

\subsection{Effective Higgs mass}
\label{sec_higgsmass}
\begin{figure}[ht]
\centering
\includegraphics[scale=0.585]{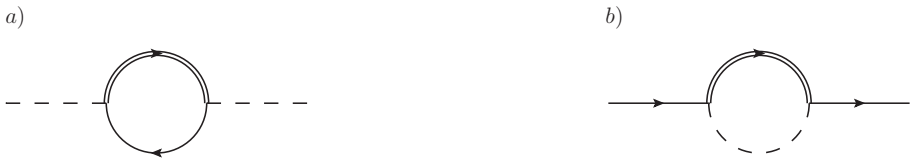}
\caption{One-loop self-energy diagrams for the Higgs, diagram $a)$, and lepton-doublet propagators, diagram $b)$, 
in the fundamental theory \eqref{Lag1}. The solid double line in the loop stands for the propagators 
of the heavier Majorana neutrinos with masses $M_i$ ($i=2,3$).}
\label{fig:self_mi}
\end{figure}

At the level of the EFT$_1$ a finite Higgs mass is generated from matching loop corrections to the Higgs propagator in the fundamental theory, 
which involve heavy Majorana neutrinos with mass $M_i$, with the EFT$_1$ operator $-m_\phi^2 \phi^\dagger \phi$.
The relevant one-loop diagram is diagram~$a)$ of figure~\ref{fig:self_mi}.
Note that, because of chiral symmetry, the one-loop correction to the lepton-doublet propagator vanishes (see diagram~$b)$ of figure~\ref{fig:self_mi}).

From the self-energy diagram $a)$ of figure~\ref{fig:self_mi} one obtains, after renormalizing in the $\overline{{\rm{MS}}}$ scheme, 
\begin{equation} 
m_\phi^2  = 2\frac{M_i^2|F_i|^2}{(4 \pi)^2}\left[1 +  \ln \left( \frac{\mu^2}{M_i^2}  \right) \right] \,.
\label{mass_Higgs_M}
\end{equation} 
A sum over the index $i$ is understood. Implications of the above formula for bounds on the heavy neutrinos masses 
and Yukawa couplings can be found in~\cite{Boyarsky:2009ix}.
The correction induced to the width and to the CP asymmetry by the finite Higgs mass is of relative order $m^2_\phi/M^2_1 \sim  |F_i|^2 M_i^2/M^2_1$, 
hence it is parametrically suppressed by two Yukawa couplings with respect to the other corrections considered in this work. 
Since we systematically neglect higher-order corrections in the Yukawa couplings, 
in the following we will also neglect the effects due to the finite Higgs-boson mass \eqref{mass_Higgs_M}.

\section{EFT$_{2}$}
\label{sec2}
In this section we compute the effective field theory EFT$_2$, which is the EFT that follows from the EFT$_1$ \eqref{Lag2} 
by integrating out degrees of freedom with energies and momenta of order~$M_1$.
By integrating out energy modes of order $M_1$,
we end up with a quantum field theory whose degrees of freedom are
non-relativistic Majorana neutrinos of type 1 and SM particles with typical energies much smaller than $M_1$. 
As regards thermal corrections to the CP asymmetry, we set our accuracy at leading order
in the $M_1/M_i$ expansion, namely we restrict to those diagrams with
the effective vertices induced by the dimension-five operators in \eqref{Lag2} (see figure~\ref{fig:eft1}) only.

To compute the Wilson coefficients of the EFT$_2$ it is necessary to match it to the EFT$_1$.
As in the case of the matching of the EFT$_1$, the temperature can be set to zero 
and one needs to compute only in-vacuum matrix elements since, according to the scale hierarchy \eqref{Hiera}, 
the matching can be performed at a scale larger than $T$.
The EFT$_2$ Lagrangian is organized as an expansion in the inverse of the lightest Majorana neutrino mass, $M_1$, 
and its expression, relevant for the Majorana neutrino decay, reads~\cite{Biondini:2013xua}
\begin{equation}
\mathcal{L}_{\text{EFT}_2}=\mathcal{L}_{\hbox{\tiny SM}} + \bar{N} \left(iv \cdot \partial +i \frac{\Gamma^{T=0}}{2} \right)N  
+ \frac{\mathcal{L}_{\hbox{\tiny N-SM}}^{(1)}}{M_1} +  \frac{\mathcal{L}_{\hbox{\tiny N-SM}}^{(3)}}{M_1^3} + \dots  \, .
\label{Lag3}
\end{equation}
The field $N$ describes the low-energy modes of the lightest Majorana neutrino.
The vector $v^{\mu}$ with $v^2=1$ identifies the reference frame.
In the following we choose the reference frame where the Majorana neutrino is at rest in the infinite mass limit;
this amounts at setting $v^\mu = (1, \vec{0})$.  
The terms $\mathcal{L}_{\hbox{\tiny N-SM}}^{(1)}$ and $\mathcal{L}_{\hbox{\tiny N-SM}}^{(3)}$ comprise dimension-five
and dimension-seven operators respectively and the dots stand for higher-order operators further suppressed in $1/M_1$. 
We do not write $\mathcal{L}_{\hbox{\tiny N-SM}}^{(2)}$ because it contains operators not contributing to thermal tadpoles~\cite{Biondini:2013xua}. 
Hence these operators do not contribute to the thermal width and CP asymmetry either.

The term $\mathcal{L}_{\hbox{\tiny N-SM}}^{(1)}$ contains just one dimension-five operator that reads~\cite{Biondini:2013xua}
\begin{equation}
\mathcal{L}_{\hbox{\tiny N-SM}}^{(1)}=a \; \bar{N} N \, \phi^{\dagger} \phi \, ,
\label{Ope_Higgs}
\end{equation} 
where $a$ is a Wilson coefficient.
Contributions to the CP asymmetry are of order $F^4$, and, at leading order, depend on the SM couplings 
$\lambda$, the Higgs self-coupling, and $g$ and $g'$, the gauge couplings of the SU(2)$_L$ and U(1)$_Y$ gauge groups respectively. 
Diagrams give a leptonic contribution, $a^{\ell}$, when cutting through a lepton line 
and an anti-leptonic contribution, $a^{\bar{\ell}}$, when cutting through an anti-lepton line. 
The diagrams and the corresponding cuts are listed and computed in appendix~\ref{sec_dim_5}. 
The calculation is close to that one carried out in the case of two heavy neutrinos 
with nearly degenerate masses in~\cite{Biondini:2015gyw}. 

In the present work, we investigate also the leading thermal effects that depend on the top Yukawa coupling, $\lambda_t$. 
These are generated by some dimension-seven operators in $\mathcal{L}_{\hbox{\tiny N-SM}}^{(3)}$. 
Despite these effects being parametrically suppressed by $(T/M_1)^2$ with respect to those induced by the operator in \eqref{Ope_Higgs},
differences in the value of the SM couplings and numerical factors may alter their relative relevance at high temperatures. 
As a reference, for $T=10^9$ GeV, the SM couplings are found to be $\lambda \approx 0.004$, 
$(2g^2+g'^2) \approx 0.824$ and $|\lambda_t|^2 \approx 0.316$~\cite{Buttazzo:2013uya,Rose15}. 
We elaborate more on this in the conclusions. 

The dimension-seven operators whose Wilson coefficients get contributions proportional to $|\lambda_t|^2$ 
are~\cite{Biondini:2013xua}\footnote{
We do not consider operators that would give rise to an interaction between the heavy-neutrino spin and the medium. 
They do not contribute to thermal tadpoles in an isotropic medium. 
We also do not consider dimension-seven operators involving gauge fields, 
since their contribution proportional to $|\lambda_t|^2$ would be subleading. 
All dimension-seven operators are listed in equation~(4.6) of~\cite{Biondini:2013xua}. 
} 
\begin{eqnarray}
&&\mathcal{L}_{\hbox{\tiny N-t}}^{(3)}=c_{3} \; \bar{N}N  \left( \bar{t}P_{L} \, v^\mu v^\nu \gamma_{\mu} \, i D_{\nu} t \right) \,,
\label{Ope_t}\\
&&\mathcal{L}_{\hbox{\tiny N-Q}}^{(3)}= c_{4} \; \bar{N}N  \left( \bar{Q}P_{R} \, v^\mu v^\nu \gamma_{\mu} \, i D_{\nu} Q \right) \, ,
\label{Ope_Q}\\
&&\mathcal{L}_{\hbox{\tiny N-L}}^{(3)}= c^{hh'}_{1c} \;  \left( \bar{N} P_{R} \, i v\cdot D L^{c}_{h'} \right) \left( \bar{L}^{c}_{h} P_{L}  N \right)  
+ c^{hh'}_{1} \left(\bar{N}P_{L} \, iv\cdot D L_{h}\right) \left(\bar{L}_{h'} P_{R} N \right)    \, ,
\label{Ope_L}
\end{eqnarray}
where $t$ is the top-quark singlet field and $Q$ is the heavy-quark SU(2) doublet. 
We note that at the order we are working here, namely at order $F^4$ in the Yukawa couplings, 
we have to distinguish the Wilson coefficients relative to the two operators in \eqref{Ope_L}. 
Indeed, they are responsible for different contributions to the neutrino thermal widths: 
the former encodes cuts on leptons whereas the second encodes cuts on anti-leptons only 
(see appendix~\ref{sec_dim_7} for details). 
On the other hand, at order $F^2$ the two operators share the same Wilson coefficient~\cite{Biondini:2013xua}.

The difference between the decay widths of the lightest Majorana neutrino into a lepton, $\ell$, 
and an anti-lepton, $\bar\ell$, with flavour $f$ can be split into a vacuum and thermal part:
\begin{equation}
\Gamma(\nu_{R,1} \to \ell_{f} + X)-\Gamma(\nu_{R,1} \to \bar{\ell}_{f}+ X )  = 
\left( \Gamma_f^{\ell,T=0}-\Gamma_f^{\bar{\ell},T=0} \right)   + \left( \Gamma_f^{\ell,T}-\Gamma_f^{\bar{\ell},T} \right) \, .
\label{def_gammaT}
\end{equation}
The in-vacuum part can be taken from appendix~\ref{AppA1} (equations \eqref{lepwidth} and \eqref{antilepwidth}).
It reads at first order in $1/M_i$
\begin{equation}
\Gamma_f^{\ell,T=0}-\Gamma_f^{\bar{\ell},T=0} = - \frac{6}{(16 \pi)^2} \frac{M_1^2}{M_i}  {\rm{Im}}\left[ (F_{1}^{*}F_{i})(F^*_{f1}F_{fi}) \right] \, .
\label{Diff_zero}
\end{equation} 
From \eqref{Ope_Higgs}-\eqref{Ope_L} the thermal part of \eqref{def_gammaT} can be written as 
\begin{equation}
\Gamma^{\ell,T}_f=\Gamma^{\ell,T}_{f,\phi}+\Gamma^{\ell,T}_{f,{\rm{fermions}}} \, ,
\label{thermalwidth}
\end{equation}
with (for $v^\mu = (1, \vec{0})$)
\begin{eqnarray}
\Gamma^{\ell,T}_{f,\phi} &=& 2 \frac{{\rm{Im}} \, a_f^{\ell}}{M_1}  \,\langle \phi^\dagger(0)\phi(0)\rangle_T \, , 
\label{Gamma_lep_higgs} \\
\Gamma^{\ell,T}_{f,{\rm{fermions}}} &=& 2 \frac{{\rm{Im}} \, c_{3,f}^{\ell}}{M_1^3} \, \langle \bar{t}(0) P_L\gamma^0 iD_0 t(0) \rangle_{T} 
+ 2 \frac{{\rm{Im}} \, c_{4,f}^{\ell}}{M_1^3}  \langle \bar{Q}(0) P_R\gamma^0 iD_0 Q(0) \rangle_{T} 
\nonumber \\
&& - \frac{{\rm{Im}} \, c_{1c,f}^{hh', \ell}}{4 M_1^3} \langle \bar{L}_{h'}(0) \gamma^0 iD_0 L_h(0) \rangle_{T}  \, ,
\label{Gamma_lep_ferm}
\end{eqnarray}
where $\langle \cdots \rangle_T$ stands for the thermal average of SM fields weighted by the SM partition function. 
Similar expressions hold for $\Gamma_f^{\bar{\ell},T}$ after replacing the leptonic contributions to the Wilson coefficients 
in \eqref{Gamma_lep_higgs} and \eqref{Gamma_lep_ferm} with the anti-leptonic ones.  

The thermal part of \eqref{def_gammaT} depends on the imaginary parts of the Wilson coefficients $a_f^{\ell}$,  $a_f^{\bar\ell}$, 
$c_{3,f}^{\ell}$,  $c^{\bar{\ell}}_{3,f}$, $c_{4,f}^{\ell}$, $c^{\bar{\ell}}_{4,f}$, $c_{1c,f}^{hh', \ell}$ and $c_{1,f}^{hh', \bar{\ell}}$
appearing in \eqref{Gamma_lep_higgs}, \eqref{Gamma_lep_ferm} and in the corresponding anti-leptonic widths.
The method to compute the imaginary parts of the Wilson coefficients has been presented 
in detail in~\cite{Biondini:2013xua,Biondini:2015gyw}, hence we recall it here only briefly. 
Four-particle two-loop diagrams in the EFT$_1$ are matched to four-particle effective vertices in the EFT$_2$. 
In the case of the dimension-five operator, one has to consider diagrams with two Higgs bosons 
and two heavy Majorana neutrinos as external legs. 
The external Higgs bosons have typical momentum $q^{\mu} \sim T$, which can be set to zero in the matching. 
The complete set of diagrams is shown and computed in appendix~\ref{sec_dim_5}. 
Leptons and anti-leptons of flavour $f$ can be put on shell by properly cutting each diagram, 
so to select the contributions to $a_f^{\ell}$ and $a_f^{\bar{\ell}}$ respectively. 
The result reads at leading order in $1/M_i$ and in the SM couplings (only terms contributing to the CP asymmetry are displayed):
\begin{equation}
{\rm{Im}} \, a_f^{\ell}=-{\rm{Im}} \, a_f^{\bar{\ell}}=\frac{3}{(16 \pi)^2} \frac{M_1}{M_i} 
\left[  8 \lambda - \frac{\left( 2g^2+g'^2\right)}{4} \right]   {\rm{Im}}\left[(F_{1}^{*}F_{i})(F^*_{f1}F_{fi}) \right] \, .
\label{match_a}
\end{equation} 
The result for anti-leptons can be obtained by substituting $F_1 \leftrightarrow F_{i}$ in the leptonic result.

The dimension-seven operators in \eqref{Ope_t}-\eqref{Ope_L} generate the leading thermal contribution to the CP asymmetry 
proportional to the top-quark Yukawa coupling, which is of relative order $|\lambda_t|^2 (T/M_1)^4$.
The list of relevant diagrams and details of the computation are given in appendix~\ref{sec_dim_7}. 
The imaginary parts of the Wilson coefficients of the dimension-seven operators at leading order in $1/M_i$ read:
\begin{eqnarray}
&&\hspace{-1cm}
{\rm{Im}} \, c_{3,f}^{\ell}=
-{\rm{Im}} \, c^{\bar{\ell}}_{3,f}=-\frac{5|\lambda_t|^2}{2(16 \pi)^2} \frac{M_1}{M_i}  {\rm{Im}}\left[ (F_{1}^{*}F_{i})(F^*_{f1}F_{fi})\right] \, ,
\label{match_ct}
\\
&&\hspace{-1cm}
{\rm{Im}} \, c_{4,f}^{\ell}=
-{\rm{Im}} \, c^{\bar{\ell}}_{4,f}=
-  \frac{5|\lambda_t|^2}{4(16 \pi)^2} \frac{M_1}{M_i} {\rm{Im}}\left[ (F_{1}^{*}F_{i})(F^*_{f1}F_{fi})\right] \, ,
\label{match_cQ} 
\\
&&\hspace{-1cm}
{\rm{Im}} \, c_{1c,f}^{hh', \ell}=  
- {\rm{Im}} \, c_{1,f}^{hh', \bar{\ell}}=  
-\frac{9 |\lambda_t|^2}{(16 \pi)^2} \frac{M_1}{M_i} {\rm{Im}} \left[(F_{f1}^* F_{fi})(F^*_{h1}F_{h'i})-(F_{f1}F_{fi}^*)(F_{h'1}F^*_{hi}) \right]  \, ,
\label{match_cL} 
\end{eqnarray} 
where we show only terms proportional to $|\lambda_t|^2$ that contribute to the CP asymmetry.
Our convention here and in the following is to label with $h$ and $h'$ the flavours of the lepton doublets in the dimension-seven operators
(these leptons belong to the thermal medium and will eventually contribute to the thermal average), 
and to label with $f$ the flavour of the lepton (anti-lepton) that appears in the final state of the Majorana neutrino decay
(this is a highly-energetic lepton contributing to the CP asymmetry).

\section{CP asymmetry at finite temperature}
\label{sec3}
\begin{figure}[ht]
\centering
\includegraphics[scale=0.5]{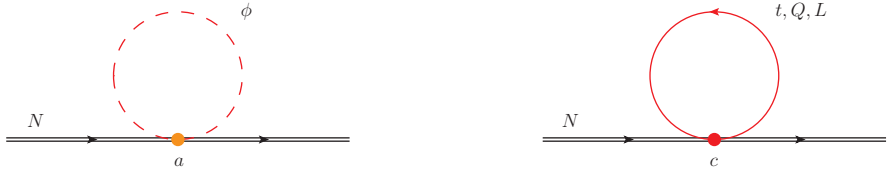}
\caption{Tadpole diagrams providing the leading thermal corrections 
to the thermal averages appearing in \eqref{Gamma_lep_higgs} and \eqref{Gamma_lep_ferm}.
Particles belonging to the thermal plasma are shown in red: 
Higgs bosons on the left, and top quarks, heavy-quark doublets and leptons on the right. 
With the vertex labeled $a$ we mean the vertex induced by the operator \eqref{Ope_Higgs}, 
whereas with the vertex labeled $c$ we mean one of the vertices induced by the operators \eqref{Ope_t}-\eqref{Ope_L}.}
\label{fig:tadpolesEFT2} 
\end{figure}

In this section, we compute the leading thermal corrections to the CP asymmetry 
proportional to the SM couplings, $\lambda$, $g^2$, $g'^2$ and $|\lambda_t|^2$.
In the framework of the EFT$_2$, thermal corrections are encoded in the thermal averages appearing in \eqref{Gamma_lep_higgs} and \eqref{Gamma_lep_ferm}.
At leading order, the thermal averages may be computed from the tadpole diagrams shown in figure~\ref{fig:tadpolesEFT2}. 
They read
\begin{eqnarray}
&&\hspace{-1cm}
\langle \phi^{\dagger}(0) \phi(0) \rangle_{T}  = \frac{T^2}{6} \, , 
\qquad\qquad\qquad\qquad 
\langle \bar{t}(0) P_L \gamma^0 iD_0 t(0) \rangle_{T}  = \frac{7 \pi^2 T^4}{40} \, ,  
\label{condensate1}\\ 
&&\hspace{-1cm}
\langle \bar{Q}(0) P_R\gamma^0   iD_0 Q(0) \rangle_{T} = \frac{7 \pi^2 T^4}{20} \, , 
\qquad 
\langle \bar{L}_{h'}(0)\gamma^0 iD_0 L_h(0) \rangle_{T} = \frac{7 \pi^2 T^4}{30} \delta_{hh'} \, .
\label{condensate2}
\end{eqnarray}
We assume the thermal bath to be at rest with respect to the lightest Majorana neutrino
and we choose the reference frame such that $v^{\mu}=(1,\vec{0})$. 

We split both the neutrino width and the CP asymmetry into a vacuum and a thermal part: 
$\Gamma=\Gamma^{T=0}+\Gamma^{T}$ and $\epsilon_f=\epsilon_f^{T=0}+\epsilon_f^{T}$. 
The decay width of the lightest Majorana neutrino reads, see~\cite{Laine:2011pq} and~\cite{Biondini:2013xua},
\begin{equation}
\Gamma = \Gamma^{T=0} + \Gamma^{T} = 
\frac{|F_1|^2M_1}{8 \pi } \left[ 1 -\lambda \left( \frac{T}{M_1}\right)^2 - \frac{7 \pi^2}{60}\,|\lambda_t|^2\,\left( \frac{T}{M_1}\right)^4 \right] \, ,
\label{widthfour}
\end{equation}
which is valid at leading order in the SM couplings in the vacuum part, $\Gamma^{T=0}= \sum_f \Gamma_f^{\ell,T=0}+\Gamma_f^{\bar{\ell},T=0}=|F_1|^2M_1/(8 \pi)$, 
and at relative order $\lambda (T/M_1)^2$ and $|\lambda_t|^2(T/M_1)^4$ in the thermal part.
In the thermal part we do not show corrections of relative order $g^2(T/M_1)^4$ and $g'^2(T/M_1)^4$ 
that are beyond the accuracy of the present work.

The in-vacuum part for the CP asymmetry, $\epsilon_f^{T=0}$, at leading order in $M_1/M_i$, can be taken from \eqref{CPhiera}.
The difference between the leptonic and anti-leptonic thermal widths defined in \eqref{thermalwidth}-\eqref{Gamma_lep_ferm} 
depends on the Wilson coefficients \eqref{match_a}-\eqref{match_cL} 
and on the thermal averages \eqref{condensate1} and \eqref{condensate2}. 
Taking them into account, it reads 
\begin{equation}
\Gamma^{\ell,T}_f-\Gamma^{\bar{\ell},T}_f = 
\frac{1}{64 \pi^2} \frac{M^2_1}{M_i}  \, {\rm{Im}}\left[ (F_{1}^{*}F_{i})(F^*_{f1}F_{fi})\right] 
\left[   \left( 4 \lambda - \frac{2g^2+g'^2}{8} \right)  \frac{T^2}{M^2_1} 
- \frac{7 \pi^2}{20}\,|\lambda_t|^2\,\left( \frac{T}{M_1}\right)^4   \right] .
\label{difftherm}
\end{equation}
Finally, from \eqref{Diff_zero}, \eqref{widthfour} and \eqref{difftherm} we obtain at order $M_1/M_i$ 
\begin{equation}
\epsilon_{f}^{T} = 
-\frac{3}{16 \pi} \frac{M_1}{M_i} \, \frac{ {\rm{Im}}\left[ (F_{1}^{*}F_{i})(F^*_{f1}F_{fi})\right]}{|F_1|^2} 
\left[   \left(  -\frac{5}{3} \lambda + \frac{2g^2+g'^2}{12} \right) \left( \frac{T}{M_1}\right)^2  
+\frac{7 \pi^2}{20}\,|\lambda_t|^2\,\left( \frac{T}{M_1}\right)^4   \right] .
\label{finalRes}
\end{equation}
This expression is valid at leading order in the SM couplings and for each SM coupling it provides the leading thermal correction. 
The thermal correction is of relative order $(T/M_1)^2$ for the Higgs self-coupling and the gauge couplings, 
and of relative order $(T/M_1)^4$ for the top Yukawa coupling. At relative order $(T/M_1)^4$ there are also corrections depending 
on the other SM couplings besides the top Yukawa coupling. Since they would provide for each coupling subleading thermal 
corrections with respect to those computed at relative order $(T/M_1)^2$, they have not been included in the present analysis.

We conclude this section by computing the leading effect to the CP asymmetry due to the Majorana neutrino motion. 
So far we have considered the neutrino at rest, for we have not included neutrino-momentum dependent operators in our list of operators.
The leading neutrino-momentum-dependent operator relevant for the neutrino decay is~\cite{Biondini:2013xua,Laine:2011pq}
\begin{equation}
\mathcal{L}_{{\hbox{\tiny N-mom.\,dep.}}} = -\frac{a}{2M_1^3} \bar{N} \left[ \partial^2-(v \cdot \partial)^2 \right] N \phi^{\dagger} \phi \, .
\label{Ope_mom}
\end{equation}
The Wilson coefficient $a$ in \eqref{Ope_mom} is the same Wilson coefficient of the dimension-five operator in \eqref{Ope_Higgs}. 
This can be inferred from the relativistic dispersion relation or using the methods of~\cite{Brambilla:2003nt}. 
When the Wilson coefficient $a$ is calculated at second order in the Yukawa couplings,
one obtains from \eqref{Ope_mom} a momentum dependent thermal correction to the total neutrino width that reads~\cite{Biondini:2013xua}
\begin{equation}
\Gamma^{T}_{\phi,{\hbox{\tiny mom.\,dep.}}}=\frac{|F_1|^2 M_1}{8 \pi} \frac{\lambda}{2} \frac{\vec{k}^2 \,T^2 }{M^4_1} \, .
\label{mom_width}
\end{equation} 
In this work, we have evaluated the CP-asymmetry relevant part of $a$ at fourth order in the Yukawa couplings.
Hence the operator in \eqref{Ope_mom} can also induce a momentum dependent asymmetry, which at leading order in the SM couplings reads 
\begin{equation}
\Gamma^{\ell,T}_{f,\phi,{\hbox{\tiny mom.\,dep.}}} - \Gamma^{\bar{\ell},T}_{f,\phi,{\hbox{\tiny mom.\,dep.}}} = 
-\frac{1}{64 \pi^2} \frac{M^2_1}{M_i} \, {\rm{Im}}\left[ (F_{1}^{*}F_{i})(F^*_{f1}F_{fi}) \right] 
\left[   \left( 2 \lambda - \frac{2g^2+g'^2}{16} \right)  \frac{\vec{k}^2 \,T^2 }{M^4_1}  \right] \, .
\label{mom_diff}
\end{equation}
or (accounting for \eqref{Diff_zero} and \eqref{mom_width})
\begin{equation}
\epsilon_{f,{\hbox{\tiny mom.\,dep.}}}^{T}=-\frac{3}{16 \pi} \frac{M_1}{M_i} \, \frac{ {\rm{Im}}\left[ (F_{1}^{*}F_{i})(F^*_{f1}F_{fi}) \right]}{|F_1|^2} 
\left(  \frac{5}{6} \lambda - \frac{2g^2+g'^2}{24} \right) \frac{\vec{k}^2 \,T^2 }{M^4_1} \, .
\label{finRes2}
\end{equation}
The parametric size of this correction depends on the Majorana neutrino thermodynamics.
For instance, if the neutrino is decoupled from the plasma, then $k \sim T$ and the relative size of \eqref{finRes2} is of order $(T/M_1)^4$, 
whereas if the neutrino is in thermal equilibrium with the plasma, then $k \sim \sqrt{M_1T}$ and the relative size of \eqref{finRes2} 
is of order $(T/M_1)^3$.

\section{Conclusions}
\label{sec4} 
In an extension of the SM that includes Majorana neutrinos heavier than the 
electroweak scale coupled to Higgs bosons and leptons through complex Yukawa couplings, 
we have computed the thermal corrections to the CP asymmetry \eqref{CPdef1} originated in the leptonic decays of 
the lightest Majorana neutrinos. We have assumed the temperature, $T$, to be smaller than 
the lightest neutrino mass, $M_1$, which in turn is much smaller than the other neutrino masses, $M_i$ $(i=2,3)$. 
Thermal corrections have been computed in terms of an expansion in the Yukawa and SM couplings, $(M_1/M_i)$ and $(T/M_1)$.
The original result of the work is in the expression of the CP asymmetry \eqref{finalRes} 
(in addition, equation \eqref{finRes2} provides the leading thermal correction depending on the Majorana neutrino momentum). 
That expression is accurate at fourth-order in the Yukawa couplings, at order $M_1/M_i$, 
at leading order in the SM couplings and for each coupling it provides the leading thermal correction. 
The present study complements an analogous recent study~\cite{Biondini:2015gyw} for the case 
of two heavy Majorana neutrinos with nearly degenerate masses relevant for resonant leptogenesis.

We perform the calculation in the flavoured regime, i.e., we assume that the flavour of the leptons and anti-leptons is resolved by the thermal medium. 
This case is relevant when the temperature at the onset of leptogenesis is smaller than $10^{12}$~GeV. 
A quantitative study of leptogenesis requires, indeed, flavour to be resolved to describe a wider range of temperatures. 
The expressions for the CP asymmetry in the unflavoured case can be recovered from \eqref{finalRes} 
(and from \eqref{finRes2}) by summing over the flavour index $f$ in the Yukawa couplings.  

The expansion in the inverse of the Majorana neutrino masses has been implemented at the Lagrangian level 
by replacing the starting theory \eqref{Lag1} with a hierarchy of two EFTs.
In the first EFT, called  EFT$_1$, energy modes of the order of the heavier neutrino masses have been integrated out.
Consequently the Lagrangian \eqref{Lag2} is organized as an expansion in $1/M_i$. 
The EFT$_1$ is characterized by operators made of two-Higgs and two-lepton fields that encompass $t$- and $s$-channel neutrino exchanges. 
Because of this, at the energy scale of the EFT$_1$, the difference between direct and indirect CP asymmetry cannot be resolved. 
We have computed the operators of dimension five and six. Operators of dimension five 
have been considered in this framework also in~\cite{Buchmuller:2000nd}. 
They contribute both to the flavoured and unflavoured CP asymmetry. 
Dimension-six operators are suppressed by $M_1/M_i$ and contribute to the flavoured CP asymmetry only. 
At the accuracy we are working, when computing thermal corrections to the CP asymmetry we neglect their contribution.
In the second EFT, called  EFT$_2$, energy modes of the order of the lightest neutrino mass have been integrated out.
The Lagrangian \eqref{Lag3} is organized as an expansion in $1/M_1$, while its dynamical degrees of freedom 
live at the energy and momentum scale of the thermal bath.
The matching of the EFT$_2$, relevant for the CP asymmetry in the leptonic decays of the lightest neutrino, is an original contribution of this work. 

\begin{figure}[ht]
\centering
\includegraphics[scale=1.1]{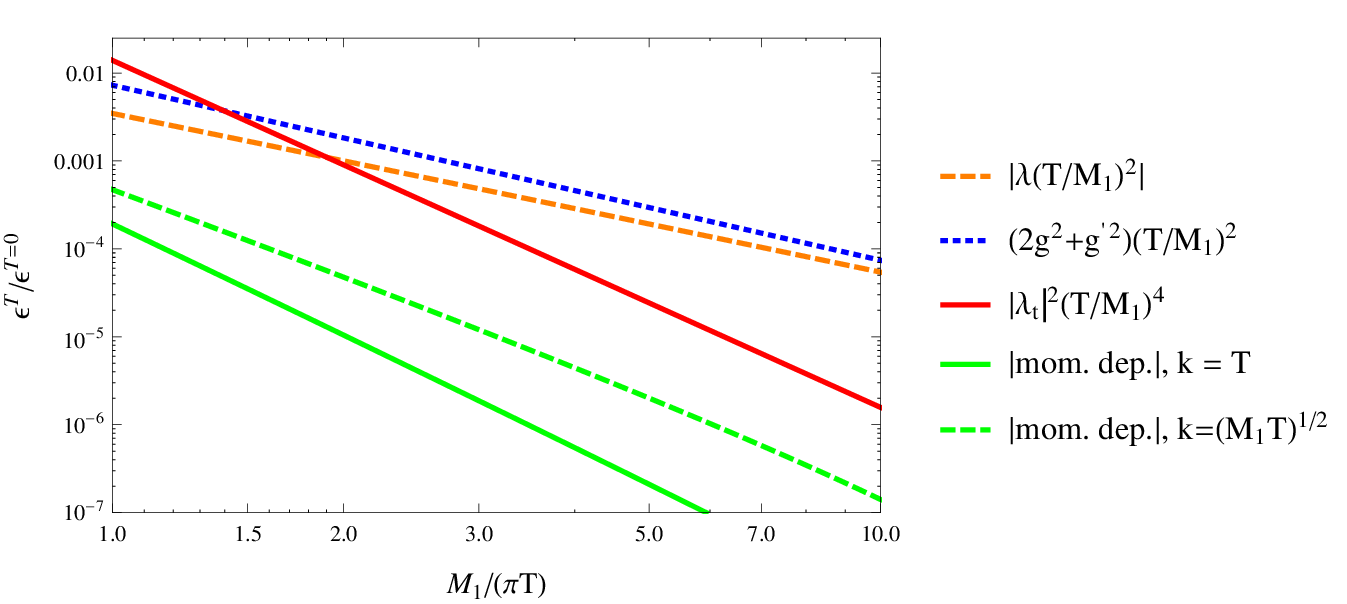}
\caption{Thermal corrections to the CP asymmetry of a Majorana neutrino decaying into leptons and anti-leptons as a function of the temperature.
The orange dashed line shows the contribution proportional to the Higgs self-coupling (the sign of the contribution 
has been changed to make it positive), the blue dotted line shows the contribution proportional to the gauge couplings 
and the red continuous line shows the contribution proportional to the top Yukawa coupling. 
These three contributions can be read from \eqref{finalRes} and refer to a neutrino at rest.
The green lines show the leading thermal contribution proportional to the neutrino momentum, which can be read from \eqref{finRes2} 
(also in this case the sign of the contribution has been changed to make it positive).
For the green continuous line we take the neutrino momentum to be $T$, whereas for the green dashed line we take it to be $\sqrt{M_1T}$.
The SM couplings have been computed at $\pi T$ with one-loop running~\cite{Rose15}. 
The different thermal contributions to the CP asymmetry have been normalized with respect to \eqref{CPhiera} at leading order in $M_1/M_i$.
The neutrino mass has been taken $M_1 = 10^7$~GeV.}
\label{fig_plotCPcontributions}
\end{figure}

Thermal corrections to the CP asymmetry in the lightest Majorana neutrino decays have been computed within the EFT$_2$. 
We have computed them at leading order in the SM couplings and for each coupling we have provided 
the leading thermal corrections, see~\eqref{finalRes}. The leading thermal corrections proportional
to the Higgs self-coupling, $\lambda$, and to the gauge couplings, $2g^2+g'^2$, are of relative order $(T/M_1)^2$, 
whereas those proportional to the top Yukawa coupling, $|\lambda_t|^2$, are of relative order $(T/M_1)^4$.
We show the different contributions in figure~\ref{fig_plotCPcontributions}. 
At low temperatures, thermal corrections proportional to the Higgs self-coupling and to the gauge couplings dominate.
However, at temperatures closer to the neutrino mass, the suppression in $T/M_1$ becomes less important 
and the numerically most relevant corrections turn out to be those proportional to the top Yukawa coupling.
In figure~\ref{fig_plotCPcontributions} we also show the thermal contribution to the CP asymmetry due to a moving Majorana neutrino, 
which has been computed in \eqref{finRes2}. We plot this contribution for the case of a neutrino with momentum $T$ 
and for the case of a neutrino in thermal equilibrium with momentum $\sqrt{M_1T}$. 
We see that for the considered momenta the effect of a moving neutrino on the thermal CP asymmetry is tiny.

\acknowledgments

We thank Miguel Angel Escobedo for collaboration on previous stages of this research project and Luigi delle Rose for communications.  
We are also grateful to Vladyslav Shta\-bo\-ven\-ko for providing useful tools to check the loop integrals~\cite{Shtabovenko:2016sxi,Vlad2}.

\appendix
\section{Matching EFT$_1$}
\label{AppA}
In this appendix we perform the tree-level matching of the operators of dimension five and six appearing in the EFT$_1$. 
To keep the notation simple, we drop the propagators on the external legs, and we re-label the so-obtained amputated Green's functions 
with the same indices used for the unamputated ones.

We start with calculating the Wilson coefficient $\eta_{ff'}^i$ of the dimension-five operator in \eqref{Lag2}. 
In order to carry out the tree-level matching, we consider the following matrix element of time-ordered operators 
in the fundamental theory \eqref{Lag1} and in the EFT$_1$ \eqref{Lag2}:
\begin{equation}
-i \int d^{4}x\,e^{i p_1 \cdot x} \int d^{4}y\,e^{i k_1 \cdot y} \int d^{4}z\,e^{i k_2 \cdot z}\, 
\langle \Omega |T( L^{\mu}_{f,m}(x) L^{\nu}_{f',n}(0) \phi_{r}(y) \phi_{s}(z) ) | \Omega \rangle \, ,
\label{matrix4part}
\end{equation}
where $\mu$ and $\nu$ are Lorentz indices, $m$, $n$, $r$ and $s$ SU(2) indices and $f,f'$ flavour indices.
When evaluating the matrix element in the fundamental theory, the result reads 
\begin{equation}
\frac{F_{fi} F_{f'i}}{M_i} (P_R C)^{\mu \nu} (\sigma^2_{mr} \sigma^2_{ns}+\sigma^2_{ms} \sigma^2_{nr} ) \, ,
\label{matrix4partbis}
\end{equation}
whereas the result in the EFT$_1$ is 
\begin{equation}
\frac{2 \eta^i_{ff'}}{M_i}(P_R C)^{\mu \nu} (\sigma^2_{mr} \sigma^2_{ns}+\sigma^2_{ms} \sigma^2_{nr} ) \, .
\label{matrix4parttris}
\end{equation}
Comparing \eqref{matrix4partbis} with \eqref{matrix4parttris}, we find the matching condition for $\eta^i_{ff'}$ given in \eqref{eq8}. 

The Wilson coefficient of the dimension-six operator in \eqref{Lag2} can be obtained in a similar fashion from the matrix element
\begin{equation}
-i \int d^{4}x\,e^{i p_1 \cdot x} \int d^{4}y\,e^{i k_1 \cdot y} \int d^{4}z\,e^{-i k_2 \cdot z}\, 
\langle \Omega |T( L^{\mu}_{f,m}(x) \bar{L}^{\nu}_{f',n}(0) \phi_{r}(y) \phi_{s}^\dagger(z) ) | \Omega \rangle \,,
\end{equation}
which computed in the fundamental theory gives 
\begin{equation}
\frac{F_{fi} F^*_{f'i}}{M^2_i} P_R^{\mu \nu} (\slashed{p}_1+\slashed{k}_1) \sigma^2_{mr} \sigma^2_{sn} \, ,
\label{matrix4ll}
\end{equation}
while computed in the EFT$_1$ is 
\begin{equation}
\frac{\tilde{\eta}^i_{ff'}}{M_i^2} P_R^{\mu \nu} (\slashed{p}_1+\slashed{k}_1) \sigma^2_{mr} \sigma^2_{sn} \, .
\label{matrix4llEFT}
\end{equation}
Comparing \eqref{matrix4ll} with \eqref{matrix4llEFT}, we find the matching condition for $\tilde{\eta}^i_{ff'}$ given in \eqref{eq8}.

\subsection{CP asymmetry at zero temperature in  the EFT$_1$}
\label{AppA1}
In the EFT$_1$ we compute now the zero temperature CP asymmetry in the leptonic decays of the lightest Majorana neutrino, $\nu_{R,1}$, 
at first and second order in $1/M_i$. To calculate the asymmetry we have to compute CP violating 
contributions to the Majorana neutrino decay widths into leptons and anti-leptons. 
This requires to compute imaginary parts of Feynman diagrams and isolate the contributions from the 
leptonic and anti-leptonic decays. Furthermore, to contribute to the CP asymmetry the diagrams must be 
sensitive to the complex phase of the Majorana neutrino Yukawa couplings.
This restricts the computation to the imaginary parts of (at least) two-loop Feynman diagrams in the EFT$_1$, 
which are of fourth order in the Majorana neutrino Yukawa couplings 
and sensitive to their complex phase through the interference of two different neutrino species, see \eqref{CPdef2}.

Given a Feynman diagram ${\mathcal{D}}$, the imaginary part ${\rm{Im}}(-i {\mathcal{D}})$ can be computed 
by means of the cutting equation~\cite{Cutkosky:1960sp,Remiddi:1981hn,Bellac,Denner:2014zga} 
\begin{equation}
{\rm{Im}}(-i \mathcal{D})= - {\rm{Re}}(\mathcal{D})=\frac{1}{2} \sum_{\rm cuts} \mathcal{D} \, ,
\label{cuttingEquation}
\end{equation} 
where the sum runs over all possible cuts of the diagram ${\mathcal{D}}$. 
If we are interested in leptonic decays, we may restrict the cuts to include leptonic lines.
Viceversa, if we are interested in anti-leptonic decays, we may restrict the cuts to include anti-leptonic lines.
We will represent a cut by a blue thick dashed line. 
The cut equation requires that vertices on the right of the cut are circled. 
Circled vertices have opposite sign than uncircled vertices. 
Propagators between uncircled vertices are the usual Feynman propagators,
propagators between circled vertices are the complex conjugate of the uncircled propagators 
and propagators between one circled and one uncircled vertex describe an on-shell particle.
By means of the cutting equation and the above cutting rules, we may isolate from two-loop diagrams 
the relevant contribution to the CP asymmetry in the leptonic and anti-leptonic decays.
A detailed description of the technique applied to the present case can be found in~\cite{Biondini:2015gyw}. 

\begin{figure}[ht] 
\centering
\includegraphics[scale=0.5]{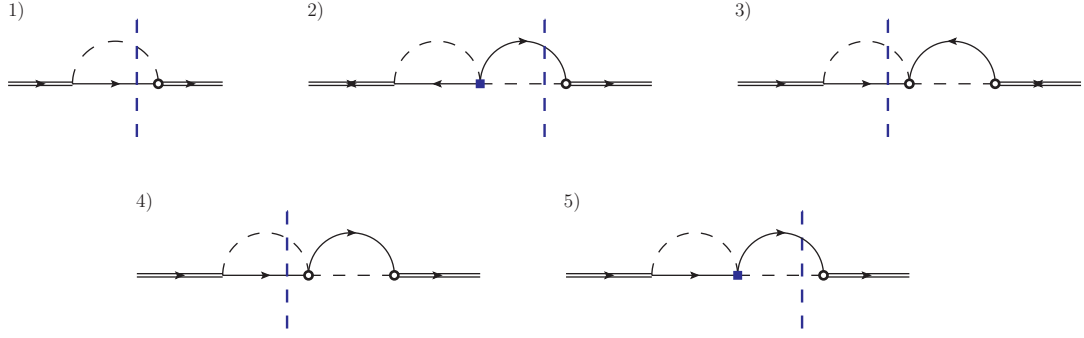}
\caption{One-loop and two-loop self-energy diagrams in the EFT$_1$ that admit cuts on a lepton line. 
Cuts are represented by vertical blue dashed lines. 
Vertices on the right of the cuts are circled; we have suppressed the symbol for the effective vertex when the vertex is circled. 
Two-Higgs-two-lepton effective vertices in the upper raw come from the dimension-five operators in \eqref{Lag2}
(vertex defined in figure~\ref{fig:eft1}), 
whereas two-Higgs-two-lepton effective vertices in the lower raw come from the dimension-six operator in \eqref{Lag2}
(vertex defined in figure~\ref{fig:eft1_bis}).}
\label{fig:appA_1}
\end{figure}

The CP violating contributions to the decay of a Majorana neutrino, $\nu_{R,1}$, 
into a lepton of flavour $f$ (and a Higgs boson) at zero temperature, whose width is $\Gamma_f^{\ell,T=0}$, 
can be computed from the imaginary parts of the diagrams shown in figure~\ref{fig:appA_1}.
The relevant leptonic cuts are also displayed.
An explicit calculation up to relative order $(M_1/M_i)^2$ gives 
\begin{eqnarray}
&&\!\!
\delta^{\mu\nu}\,\frac{\Gamma_f^{\ell,T=0}}{2} = 
{\rm Im}\left[-i \sum_{n=1}^5(\mathcal{D}^{\ell}_{n,\hbox{\tiny fig.\ref{fig:appA_1}}})\right]  = 
\label{lepwidth}
\\
&&\!\!
 \delta^{\mu \nu} \frac{M_1}{16 \pi}\left\lbrace  \frac{|F_{f1}|^2}{2}  
- 3  \frac{M_1}{M_i} \,  \frac{ {\rm{Im}}\left[(F_{1}^{*}F_{i})(F^*_{f1}F_{fi}) \right] }{32 \pi}    
- 2  \left( \frac{M_1}{M_i}\right)^2 \,  \frac{ {\rm{Im}}\left[(F_{1}F_{i}^{*})(F^*_{f1}F_{fi}) \right] }{32 \pi} + \dots \right\rbrace ,
\nonumber 
\end{eqnarray}
where the dots stand both for terms that do not contribute to the CP asymmetry at fourth-order in the Yukawa couplings 
(e.g., the real part of the Yukawa-coupling combination), and for higher-order terms in the $M_1/M_i$ expansion.  
The term proportional to $|F_{f1}|^2$, which comes from the one-loop diagram, contributes only to the leptonic width, but not to the CP asymmetry.

\begin{figure}[ht]
\centering
\includegraphics[scale=0.5]{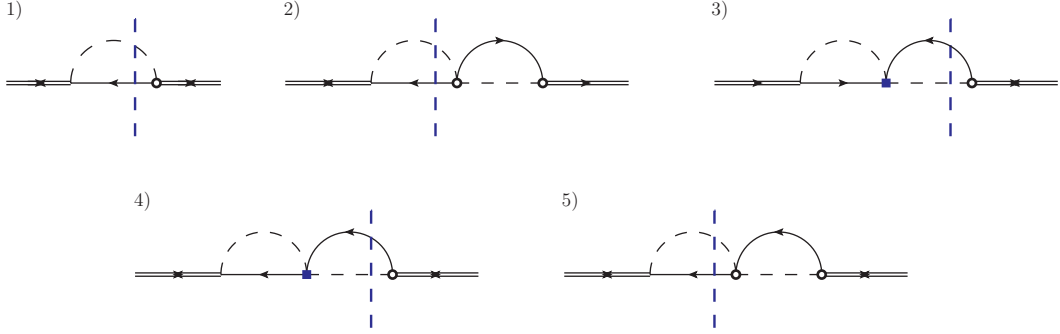}
\caption{One-loop and two-loop self-energy diagrams in the EFT$_1$ that admit cuts on an anti-lepton line. 
Further explanations and comments are like in figure~\ref{fig:appA_1}.}
\label{fig:appA_1_2}
\end{figure}

In order to compute the decay width into anti-leptons we have to consider the diagrams and the corresponding cuts on anti-lepton lines 
shown in figure~\ref{fig:appA_1_2}. The calculation up to relative order $(M_1/M_i)^2$ gives
\begin{eqnarray}
&& \!\!
\delta^{\mu\nu}\,\frac{\Gamma_{f}^{\bar{\ell},T=0}}{2} = 
{\rm Im}\left[-i \sum_{n=1}^5(\mathcal{D}^{\bar{\ell}}_{n,\hbox{\tiny fig.\ref{fig:appA_1_2}}})\right]  = 
\label{antilepwidth}  
\\
&& \!\!
\delta^{\mu \nu} \frac{M_1}{16 \pi}\left\lbrace  \frac{|F_{f1}|^2}{2}  + 3  \frac{M_1}{M_i} \,  
\frac{ {\rm{Im}}\left[(F_{1}^{*}F_{i})(F^*_{f1}F_{fi}) \right] }{32 \pi}    + 2  \left( \frac{M_1}{M_i}\right)^2 \,  
\frac{ {\rm{Im}}\left[(F_{1}F_{i}^{*})(F^*_{f1}F_{fi}) \right] }{32 \pi} + \dots \right\rbrace ,
\nonumber 
\end{eqnarray}
where the only difference with respect to \eqref{lepwidth} is in a change of sign for each coefficient with four Yukawa couplings.
The one-loop diagram contributes only to the anti-leptonic width, but not to the CP asymmetry.

Hence the CP asymmetry at zero temperature, as defined in \eqref{CPdef1}, reads at leading order 
in the SM couplings, at fourth-order in the Yukawa couplings 
and up to relative order $(M_1/M_i)^2$ in the heavy Majorana neutrino mass expansion
\begin{eqnarray}
&&\!\!
\epsilon^{T=0}_{f}=\frac{\Gamma^{\ell,T=0}_f - \Gamma^{\bar{\ell},T=0}_f}{\sum_f \Gamma^{\ell,T=0}_f + \Gamma^{\bar{\ell},T=0}_f}=
\nonumber \\
&& \!\!
-\frac{3}{16 \pi} \frac{M_1}{M_i}  \frac{{\rm{Im}} \left[ (F_{1}^{*}F_{i})(F^*_{f1}F_{fi})\right] }{|F_1|^2}  
- \frac{1}{8 \pi} \left( \frac{M_1}{M_i}\right)^2  \frac{{\rm{Im}} \left[ (F_{1}F_{i}^{*})(F^*_{f1}F_{fi})\right] }{|F_1|^2}\, .
\label{CPhiera}
\end{eqnarray}
The result coincides with the sum of the direct and indirect contributions obtained in the hierarchical limit 
from the general expressions given in \eqref{CPvert} and \eqref{CPself}.

\section{Matching EFT$_2$}
\label{AppB}
In this appendix we compute the Wilson coefficients \eqref{match_a}-\eqref{match_cL} of the EFT$_2$.
The Wilson coefficients are obtained by matching four-point Green's functions calculated in the EFT$_1$ 
with four-point Green's functions calculated in the EFT$_2$. 
Since we are going to consider effects that are of first order in the SM couplings, 
we need to specify the SM Lagrangian, which reads
\begin{eqnarray}
\mathcal{L}_{\hbox{\tiny {SM}}} &=&
\bar{L}_{f} P_R\, i \slashed{D} \, L_{f} + \bar{Q}P_R\, i\slashed{D} \, Q 
+ \bar{t}P_L\, i\slashed{D} \, t -\frac{1}{4}W_{\mu\nu}^aW^{a\,\mu\nu} -\frac{1}{4}F_{\mu\nu}F^{\mu\nu}
\nonumber\\
&& + \left( D_{\mu} \phi \right)^{\dagger}\left( D^{\mu} \phi \right)  - \lambda \left( \phi^{\dagger}\phi \right)^2 
- \lambda_{t}    \, \bar{Q} \, \tilde{\phi} \, P_{R} t
- \lambda^{*}_{t} \, \bar{t} P_{L} \, \tilde{\phi}^{\dagger} \, Q  + \dots \,,
\label{SMlag}
\end{eqnarray}
where the dots stand for terms irrelevant for our calculation. 
The Lagrangian exhibits an unbroken SU(2)$_L \times$ U(1)$_Y$ gauge
symmetry, according to the assumption $T \gg M_W$. 
The covariant derivative in \eqref{SMlag} reads, when acting on left-handed doublets 
(only the coupling with $B_\mu$ has to be considered for right-handed fermions)
\begin{equation}
D_{\mu}=\partial_{\mu} -igA^{a}_{\mu} \tau^{a} -ig'YB_{\mu} \, ,
\label{SMCov}
\end{equation}
where $\tau^{a}$ are the SU(2)$_L$ generators and $Y$ is the hypercharge ($Y=1/2$ for the Higgs boson, 
$Y=-1/2$ for left-handed leptons).  The couplings $g$, $g'$, $\lambda$ and $\lambda_t$ are the
SU(2)$_L$ and U(1)$_Y$ gauge couplings, the Higgs self-coupling and the top Yukawa coupling respectively.  
The fields $L_{f}$ are the SU(2)$_L$ lepton doublets with flavour $f$, 
$Q^T=(t,b)$ is the heavy-quark SU(2)$_L$ doublet, $A^{a}_{\mu}$ are the SU(2)$_L$ gauge fields, 
$B_{\mu}$ the U(1)$_Y$ gauge fields and $W^{a\,\mu\nu}$, $F_{\mu\nu}$ the corresponding field strength tensors, 
$\phi$ is the Higgs doublet, $t$ is the top quark field.

As mentioned in the main body of the paper, when matching EFT$_2$ with EFT$_1$ we can set the temperature to zero. 
This comes from the fact that we integrate out only high-energy modes of order $M_1 \gg T$.
Dimensional regularization is used throughout all calculations.
As a consequence all loop diagrams in the EFT$_2$ side of the matching are scaleless, and therefore vanish in dimensional regularization. 
The operators that we match are the dimension-five operator \eqref{Ope_Higgs} and the dimension-seven operators \eqref{Ope_t}-\eqref{Ope_L}
(of which we consider only the top Yukawa coupling contributions).
Therefore we need to consider matrix elements with two external heavy neutrinos and two
external Higgs bosons, two external top-quarks, two external heavy-quark doublets and two external lepton doublets.

\subsection{Matching the dimension-five operator}
\label{sec_dim_5}
In order to determine the CP violating contributions to the Wilson coefficient of the dimension-five operator of the EFT$_2$, 
we consider the following matrix element in the Majorana neutrino rest frame 
\begin{equation}
-i \left.\int d^{4}x\,e^{i p \cdot x} \int d^{4}y \int d^{4}z\,e^{i q \cdot (y-z)}\, 
\langle \Omega | T(\psi^{\mu}_{1}(x) \bar{\psi}^{\nu }_{1}(0) \phi_{m}(y) \phi_{n}^{\dagger}(z) )| \Omega \rangle
\right|_{p^\alpha =(M_1 + i\epsilon,\vec{0}\,)},
\label{B1}
\end{equation} 
where $\mu$ and $\nu$ are Lorentz indices, and $m$ and $n$ are SU(2) indices.  
The matrix element \eqref{B1} can be understood as a $2 \rightarrow 2$ scattering in the EFT$_1$ between a heavy Majorana neutrino at rest
and a Higgs boson carrying momentum $q^\mu$ much smaller than $M_1$ that can be eventually set to zero.
We divide the calculation as follows. 
First, we compute Feynman diagrams involving the Higgs self-coupling, $\lambda$, 
and, then, we compute Feynman diagrams with gauge bosons.

\begin{figure}[ht]
\centering
\includegraphics[scale=0.55]{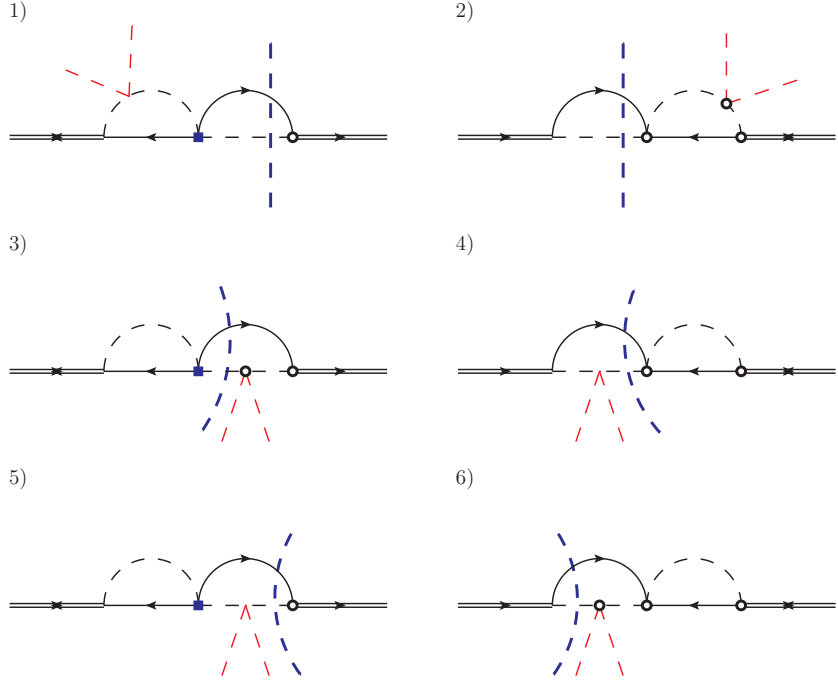}
\caption{First set of diagrams involving the Higgs self-coupling with cuts on the lepton lines.
Vertices on the right of the cuts are circled. External Higgs bosons are in red.}
\label{fig:Higgsmatch} 
\end{figure}

\begin{figure}[ht]
\centering
\includegraphics[scale=0.55]{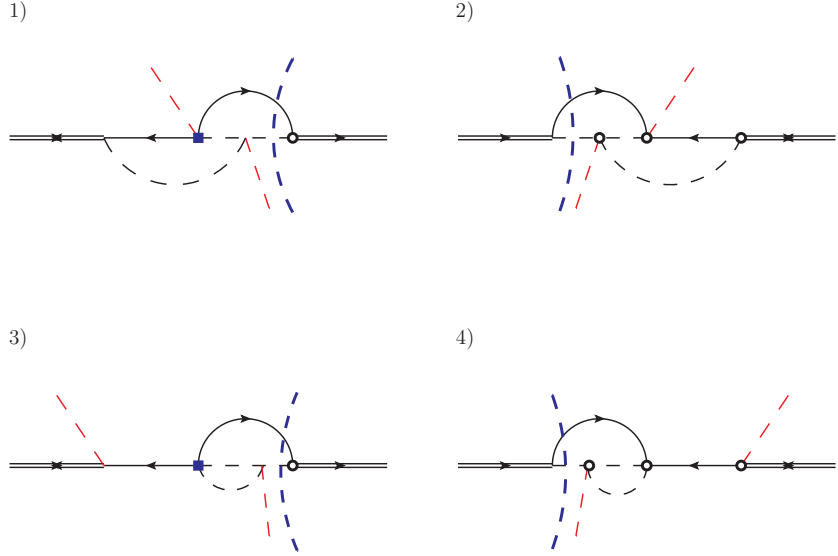}
\caption{Second set of diagrams involving the Higgs self-coupling with cuts on the lepton lines.}
\label{fig:Higgsmatchbis} 
\end{figure}

In figure~\ref{fig:Higgsmatch} and~\ref{fig:Higgsmatchbis} we list the diagrams contributing to the Wilson coefficient 
of the dimension-five operator that involve the Higgs self-coupling. 
In each raw we show a diagram and its complex conjugate and we draw explicitly the cuts that put a lepton on shell (dashed blue line).  
The diagrams in figure~\ref{fig:Higgsmatch} are obtained by adding a four-Higgs vertex to the diagrams $a)$ and $b)$ in figure~\ref{fig:eft1cp}.  
On the other hand, one can also open up one of the Higgs propagators in those diagrams, 
keep one Higgs line as an external line and connect the other one to a four-Higgs vertex added 
to the remaining internal Higgs line. These diagrams are shown in figure~\ref{fig:Higgsmatchbis}.

Starting from the diagrams in figure~\ref{fig:Higgsmatch}, we obtain 
\begin{eqnarray}
&& \!\!\!\!
{\rm{Im}}\, (-i \mathcal{D}^{\ell}_{1,\hbox{\tiny fig.\ref{fig:Higgsmatch}}})+{\rm{Im}}\, (-i \mathcal{D}^{\ell}_{2,\hbox{\tiny fig.\ref{fig:Higgsmatch}}}) =
\frac{\lambda}{M_i}\frac{9 }{(16 \pi)^2}  {\rm{Im}}\left[ (F_{1}^{*}F_{i})(F^*_{f1}F_{fi}) \right] \delta^{\mu \nu}  \delta_{mn} 
 + \dots \,, 
\label{B2} \\
&&\!\!\!\!
{\rm{Im}}\, (-i \mathcal{D}^{\ell}_{3,\hbox{\tiny fig.\ref{fig:Higgsmatch}}})+ {\rm{Im}}\, (-i \mathcal{D}^{\ell}_{4,\hbox{\tiny fig.\ref{fig:Higgsmatch}}}) 
+{\rm{Im}}\, (-i \mathcal{D}^{\ell}_{5,\hbox{\tiny fig.\ref{fig:Higgsmatch}}})+ {\rm{Im}}\, (-i \mathcal{D}^{\ell}_{6,\hbox{\tiny fig.\ref{fig:Higgsmatch}}})  = 
\nonumber \\
&&\hspace{3.6 cm}\frac{\lambda}{M_i}\frac{9 }{(16 \pi)^2}  {\rm{Im}}\left[(F_{1}^{*}F_{i})(F^*_{f1}F_{fi}) \right] \delta^{\mu \nu}  \delta_{mn}   
+ \dots \,,
\label{B3}
\end{eqnarray}
where the subscripts of $\mathcal{D}$ refer to the diagrams as listed in figure~\ref{fig:Higgsmatch} 
and the superscript, $\ell$, stands for leptonic contributions only.  
The dots in \eqref{B2} and \eqref{B3} stand for terms that are of higher
order in the neutrino mass expansion and for terms that cancel in the calculation of the CP asymmetry. 
The result for the anti-leptonic contributions differs for an overall minus sign, 
and may be obtained by replacing $F_1 \leftrightarrow F_i$ in the above expressions.
  
The diagrams shown in figure~\ref{fig:Higgsmatchbis} give 
\begin{eqnarray}
&&\hspace{-1cm}
{\rm{Im}}\, (-i \mathcal{D}^{\ell}_{1,\hbox{\tiny fig.\ref{fig:Higgsmatchbis}}})+ {\rm{Im}}\, (-i \mathcal{D}^{\ell}_{2,\hbox{\tiny fig.\ref{fig:Higgsmatchbis}}})  = 
\frac{\lambda}{M_i}\frac{6 }{(16 \pi)^2}  {\rm{Im}}\left[ (F_{1}^{*}F_{i})(F^*_{f1}F_{fi}) \right] \delta^{\mu \nu}  \delta_{mn}   
+ \dots \,, 
\label{B4}
\\
&&\hspace{-1cm}
{\rm{Im}}\, (-i \mathcal{D}^{\ell}_{3,\hbox{\tiny fig.\ref{fig:Higgsmatchbis}}})
+ {\rm{Im}}\, (-i \mathcal{D}^{\ell}_{4,\hbox{\tiny fig.\ref{fig:Higgsmatchbis}}})  = 0 \, .
\label{B6}
\end{eqnarray}
We can understand the result in \eqref{B6} as follows. 
After the cut on the lepton line the remaining loop amplitude gives a vanishing imaginary part, 
what we called ${\rm{Im}(B)}$ in \eqref{CPdef2}. 
Indeed, as we noticed in an analogous situation in~\cite{Biondini:2015gyw}, 
the momentum of the external Higgs boson can be put to zero and hence, 
after cutting the remaining loop amplitude to get the imaginary part,  
we have three on-shell massless particles entering the same vertex. 
In such a case the available phase space vanishes in dimensional regularization.

\begin{figure}[ht]
\centering
\includegraphics[scale=0.48]{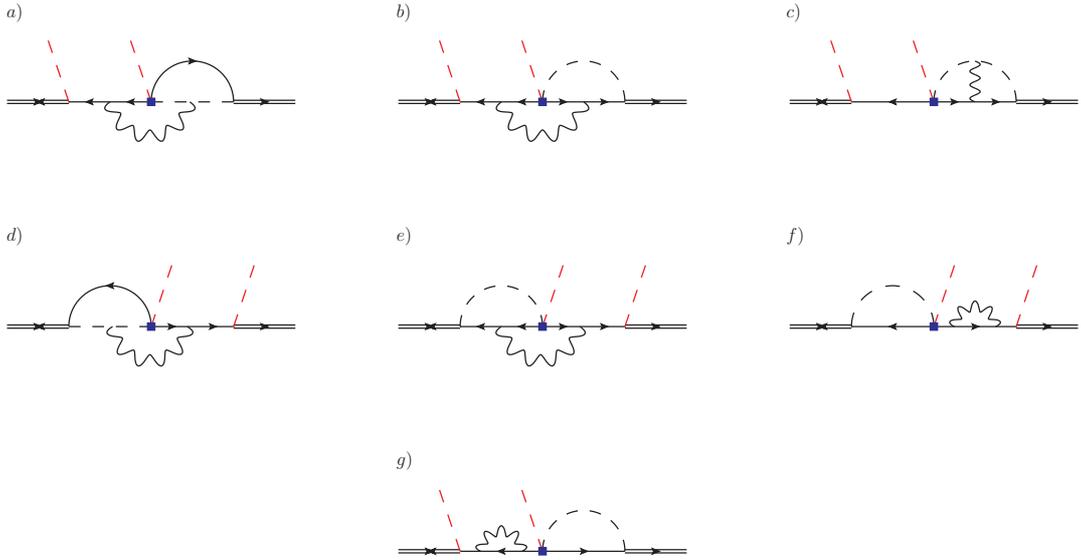}
\caption{Diagrams of order $(F_1^*F_i)(F_{f1}^*F_{fi})$ and leading order in the gauge couplings relevant for the CP asymmetry 
in the matching of $a^{\ell}_f$ and $a^{\bar{\ell}}_f$ (complex conjugate diagrams are not displayed). }
\label{fig:fig_gauge_set1} 
\end{figure}

\begin{figure}[ht]
\centering
\includegraphics[scale=0.45]{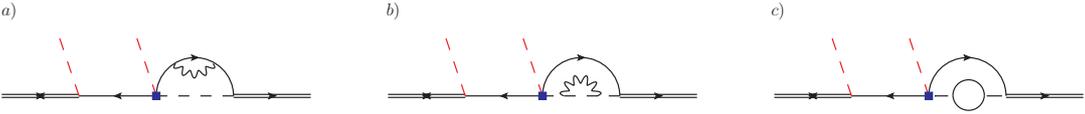}
\caption{Some diagrams of order $(F_1^*F_i)(F_{f1}^*F_{fi})$ and first order in the gauge couplings 
that do not contribute to the CP asymmetry. 
Diagram $c)$ contains a top-quark-heavy-quark-doublet loop instead of a gauge boson
(it would be a potential contribution of order $|\lambda_t|^2$ to the operator of dimension five), 
but falls in the same topology class as the first two diagrams and does not contribute to the CP asymmetry.
Complex conjugate diagrams are not displayed.}
\label{fig:noCP} 
\end{figure}

We consider now Feynman diagrams with gauge bosons.
They contribute to the Wilson coefficient of the dimension-five operator, 
and provide a dependence on the couplings of the unbroken SU(2)$_{L}$ and U(1)$_{Y}$ gauge groups, $g$ and $g'$ respectively. 
The topologies of the diagrams that could potentially contribute to the CP asymmetric part of the Wilson coefficients $a^{\ell}_f$ and $a^{\bar{\ell}}_f$
are shown in figures~\ref{fig:fig_gauge_set1} and~\ref{fig:noCP}.
We have discussed extensively how to address the calculation of diagrams involving the gauge bosons in~\cite{Biondini:2015gyw}, 
and, for this reason, we recall here only the main outcomes.

To perform calculations with gauge bosons we need to fix a gauge.
We can distinguish two different cases when cutting a lepton line
in the diagrams of figure~\ref{fig:fig_gauge_set1}: 
first, a lepton is cut with a Higgs boson, second, a lepton is cut with a gauge boson. 
These different cuts correspond to different physical processes, one without and one with gauge bosons in the final states, 
that we can treat within different gauges.

We adopt the Landau gauge for diagrams in which the lepton is cut together with a Higgs boson (the gauge boson is uncut), 
while we use the Coulomb gauge when a gauge boson is cut. 
According to this choice, we can neglect all the diagrams with a gauge boson attached to an external Higgs boson leg. 
Indeed, the vertex interaction between a gauge and a Higgs boson is proportional to the momentum of the latter
both in Landau and Coulomb gauge (see \eqref{SMlag} and \eqref{SMCov}). 
If it depends on the external momentum, it will contribute to the matching of higher-order operators 
containing derivatives acting on the Higgs fields. 
On the other hand, if it depends on the internal momentum then its contraction with the propagator vanishes
both in Coulomb gauge, if the gauge boson is cut, and in Landau gauge if the gauge boson is uncut. 
Note that in Coulomb gauge only transverse gauge bosons can be cut.

Diagram~$c)$ in figure~\ref{fig:fig_gauge_set1} is similar to one diagram, diagram $c)$ of figure~15 in~\cite{Biondini:2015gyw},
studied in the case of nearly degenerate neutrino masses and vanishes for the same reason.
The diagram may be cut in two different ways in order to put on shell a lepton together with a Higgs boson. 
The only difference between the imaginary parts of the remaining one-loop subdiagrams is in the number of circled vertices 
that leads to two contributions with opposite signs eventually cancelling each other. 
Diagram~$g)$ contains a sub-diagram that vanishes in Landau gauge
after having cut the Higgs and lepton lines, see diagram 5) in figure~4 and equation (A.8) in~\cite{Biondini:2013xua}.

The three diagrams in figure~\ref{fig:noCP} do not develop an imaginary part for the remaining loop amplitude, ${\rm{Im}}(B)$ in \eqref{CPdef2}, 
after having cut the lepton line. 
This has also been discussed in the case of nearly degenerate neutrino masses in~\cite{Biondini:2015gyw}. 
The different heavy-neutrino mass arrangement does not change the argument. 
Let us consider, for instance, diagram~$a)$ in figure~\ref{fig:noCP}, 
and let us cut it in all possible ways that put a lepton on shell. 
A first cut through the gauge boson separates the diagram into tree-level sub-diagrams. 
Since there is no loop uncut, we cannot generate any additional complex phase. 
A second and third cut are such to leave an uncut one-loop sub-diagram.
However no additional phase is generated by this sub-diagram either. 
The incoming and outgoing particles are on shell and massless, and the particles in the loop are massless as well. 
The imaginary part of the sub-diagram corresponds to a process in which three massless particles enter the same vertex, 
whose available phase space vanishes in dimensional regularization. 
Therefore the diagrams in figure~\ref{fig:noCP} can give rise only to terms proportional to 
${\rm{Re}}\left[(F_{1}^{*}F_{i})(F^*_{f1}F_{fi}) \right]$ that do not contribute to the CP asymmetry.

\begin{figure}[ht]
\centering
\includegraphics[scale=0.55]{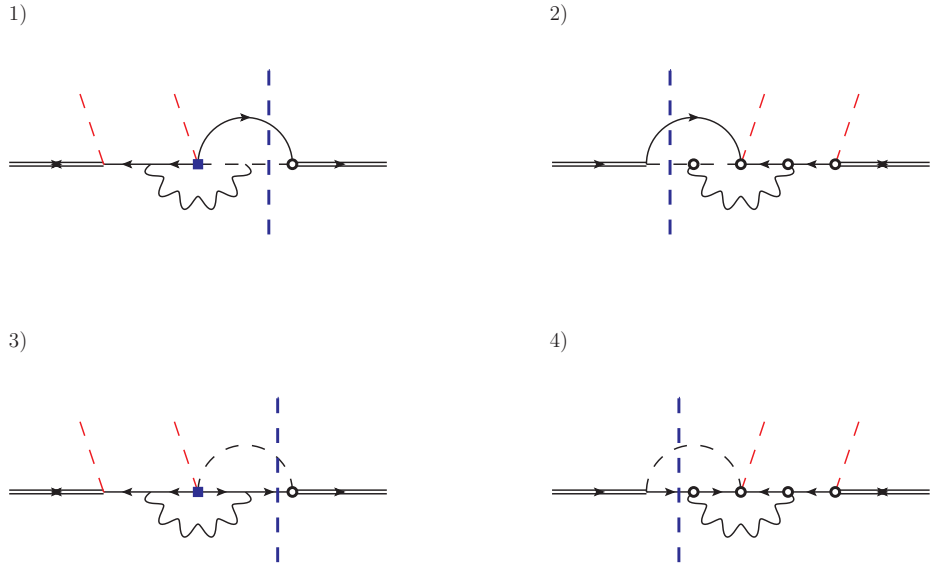}
\caption{In each raw we draw the diagrams $a)$ and $b)$ of figure~\ref{fig:fig_gauge_set1} together with 
their complex conjugate when a lepton propagator is cut with a Higgs boson propagator.}
\label{fig:leptohiggs} 
\end{figure}

\begin{figure}[ht]
\centering
\includegraphics[scale=0.52]{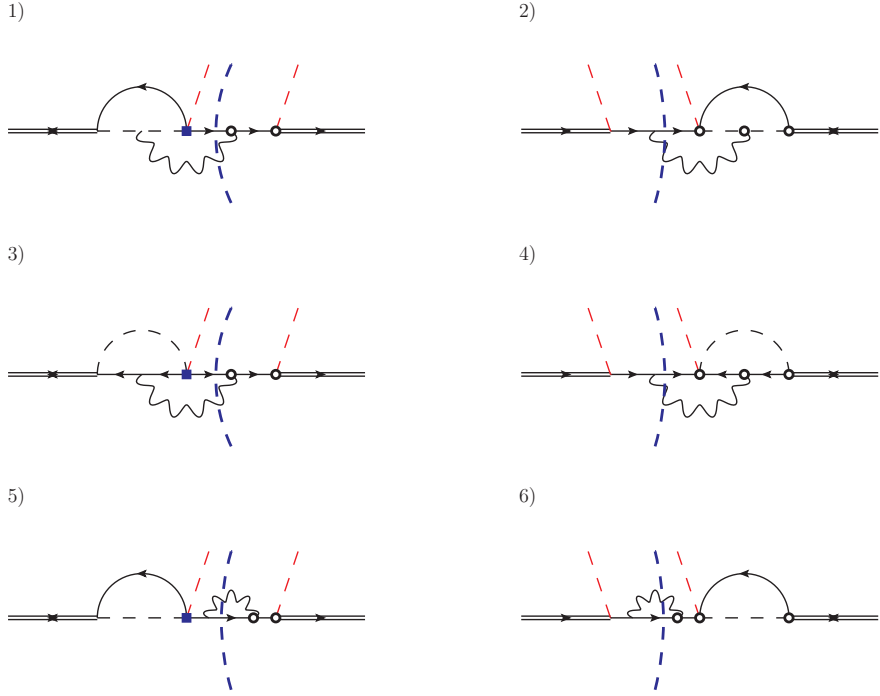}
\caption{In each raw we draw the diagrams $d)$, $e)$ and $f)$ of figure~\ref{fig:fig_gauge_set1} together with 
their complex conjugate when a lepton propagator is cut with a gauge boson propagator.} 
\label{fig:leptogauge}
\end{figure}

We finally compute the diagrams that are not excluded by the above arguments. 
They are shown in figure~\ref{fig:leptohiggs} and~\ref{fig:leptogauge}, 
where the lepton line is cut together with a Higgs boson or a gauge boson respectively. 
In each raw we show a diagram and its complex conjugate. 
We start with the diagrams in figure~\ref{fig:leptohiggs} and we recall that they are computed in Landau gauge.  
The results read
\begin{eqnarray}
&&  \hspace{-1cm}
{\rm{Im}}\, (-i \mathcal{D}^{\ell}_{1,\hbox{\tiny fig.\ref{fig:leptohiggs}}})+{\rm{Im}}\, (-i \mathcal{D}^{\ell}_{2,\hbox{\tiny fig.\ref{fig:leptohiggs}}}) = 0 , 
 \label{B7}  \\
&&  \hspace{-1cm}
{\rm{Im}}\,  (-i \mathcal{D}^{\ell}_{3,\hbox{\tiny fig.\ref{fig:leptohiggs}}})+{\rm{Im}}\, (-i \mathcal{D}^{\ell}_{4,\hbox{\tiny fig.\ref{fig:leptohiggs}}})= 
-3\left( g^2+g'^2\right) \frac{{\rm{Im}}\left[ (F_{1}^{*}F_{i})(F^*_{f1}F_{fi}) \right]}{8(16 \pi)^2 \, M_i}    \delta^{\mu \nu}  \delta_{mn}
  + \dots ,  
\label{B8}
\end{eqnarray}
where the superscript $\ell$ stands for leptonic contribution and the subscript refers to the diagram label as listed in figure~\ref{fig:leptohiggs}. 
The dots stand for higher-order terms in the heavy-neutrino mass expansion and for terms that do not contribute to the CP asymmetry.

The diagrams in figure~\ref{fig:leptogauge}, where a gauge boson appears in the final state, are computed in Coulomb gauge.
The results read 
\begin{eqnarray}
&& \hspace{-1cm}
{\rm{Im}}\, (-i \mathcal{D}^{\ell}_{1,\hbox{\tiny fig.\ref{fig:leptogauge}}})+{\rm{Im}}\, (-i \mathcal{D}^{\ell}_{2,\hbox{\tiny fig.\ref{fig:leptogauge}}}) = 
3 \left( g^2+g'^2\right)  \frac{{\rm{Im}}\left[(F_{1}^{*}F_{i})(F^*_{f1}F_{fi})\right] }{8(16 \pi)^2 M_i} \delta^{\mu \nu}  \delta_{mn} + \dots , 
\label{B9} \\
&& \hspace{-1cm}
{\rm{Im}}\,  (-i \mathcal{D}^{\ell}_{3,\hbox{\tiny fig.\ref{fig:leptogauge}}})+{\rm{Im}}\, (-i \mathcal{D}^{\ell}_{4,\hbox{\tiny fig.\ref{fig:leptogauge}}})= 
-3 \left( g^2+g'^2\right)  \frac{{\rm{Im}}\left[(F_{1}^{*}F_{i})(F^*_{f1}F_{fi})\right] }{8(16 \pi)^2 M_i} \delta^{\mu \nu}  \delta_{mn} + \dots ,  
\label{B10} \\
&& \hspace{-1cm}
{\rm{Im}}\,  (-i \mathcal{D}^{\ell}_{5,\hbox{\tiny fig.\ref{fig:leptogauge}}})+{\rm{Im}}\, (-i \mathcal{D}^{\ell}_{6,\hbox{\tiny fig.\ref{fig:leptogauge}}})=
-3 \left( 3g^2+g'^2\right)  \frac{{\rm{Im}}\left[(F_{1}^{*}F_{i})(F^*_{f1}F_{fi})\right] }{8(16 \pi)^2 M_i} \delta^{\mu \nu}  \delta_{mn} + \dots ,
\nonumber\\
\label{B11}
\end{eqnarray}
where again the superscript $\ell$ stands for leptonic contribution 
and the subscripts refer to the diagram label as listed in figure~\ref{fig:leptogauge}. 

The Wilson coefficient of the dimension-five operator can now be computed. 
In the EFT$_2$ the matrix element \eqref{B1} reads, isolating the contribution from the Majorana neutrino decaying into a lepton of flavour $f$, 
\begin{equation}
\frac{{\rm{Im}} \, a^{\ell}_f}{M_1}\delta^{\mu \nu}  \delta_{mn} \, .
\label{B12}
\end{equation}
An analogous expression holds for the decay into an anti-lepton.
Summing up \eqref{B2}-\eqref{B11} and matching with \eqref{B12}, we obtain \eqref{match_a}.

\subsection{Matching dimension-seven operators proportional to $|\lambda_t|^2$}
\label{sec_dim_7}
Here we compute the CP-violating contributions to the dimension-seven operators of the EFT$_2$ proportional to $|\lambda_t|^2$.
We will first match the operators \eqref{Ope_t} and \eqref{Ope_Q}, and then the operators~\eqref{Ope_L}.

\begin{figure}[ht]
\centering
\includegraphics[scale=0.52]{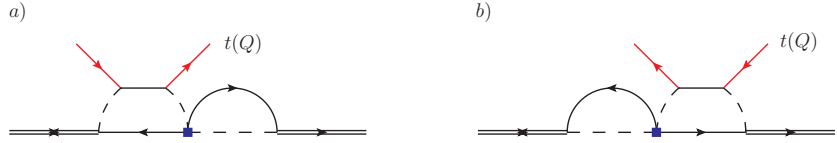}
\caption{Diagrams $a)$ and $b)$ are the two diagrams contributing to the heavy-neutrino-top quark (heavy quark doublet) operator 
with the combination of Yukawa couplings $(F_1^*F_i)(F_{f1}^*F_{fi})$ (complex conjugate diagrams are not displayed).
Top (heavy-quark doublet) external legs are shown as solid red lines. 
We drop the arrow for the internal heavy-quark doublet (top quark) in order to avoid confusion with lepton lines (arrows kept).}
\label{fig:matchtop} 
\end{figure}

A quite limited number of diagrams allows to completely specify the CP violating terms in the Wilson coefficient 
of the heavy-neutrino-top-quark (heavy-quark doublet) operator. We show them in figure~\ref{fig:matchtop}. 
The external fermion legs have to be understood as top quarks or heavy-quark doublets, as explicitly indicated.  

We consider the following matrix elements in the  EFT$_1$:
\begin{eqnarray}
&&\hspace{-10mm}
-i \left.\int d^{4}x\,e^{i p \cdot x} \int d^{4}y \int d^{4}z\,e^{i q \cdot (y-z)}\, 
\langle \Omega | T(\psi^{\mu}(x) \bar{\psi}^{\nu }(0) \, t^{\sigma}(y) \bar{t}^{\lambda}(z)) | \Omega \rangle 
\right|_{p^\alpha =(M_1 + i\epsilon,\vec{0}\,)} ,
\label{C0}\\
&&\hspace{-10mm}
-i \left.\int d^{4}x\,e^{i p \cdot x} \int d^{4}y \int d^{4}z\,e^{i q \cdot (y-z)}\, 
\langle \Omega | T(\psi^{\mu}(x) \bar{\psi}^{\nu }(0) \, Q_{m}^{\sigma}(y) \bar{Q}_{n}^{\lambda}(z)) | \Omega \rangle 
\right|_{p^\alpha =(M_1 + i\epsilon,\vec{0}\,)} .
\nonumber\\
\label{C1}
\end{eqnarray}
They describe respectively a $2 \rightarrow 2$ scattering between a heavy Majorana neutrino at rest 
and a right-handed top quark carrying momentum $q^\mu$, and a $2 \rightarrow 2$ scattering 
between a heavy Majorana neutrino at rest and a left-handed heavy-quark doublet carrying momentum $q^\mu$. 
The indices $\mu$, $\nu$, $\sigma$ and $\lambda$, are Lorentz indices, and $m$ and $n$ are the SU(2) indices of the heavy-quark doublet. 
Differently from the former matching of the dimension-five operator, 
the external momentum of the SM particles cannot be put to zero in the following calculation, 
since we match operators with derivatives acting on the external fields.

\begin{figure}[ht]
\centering
\includegraphics[scale=0.52]{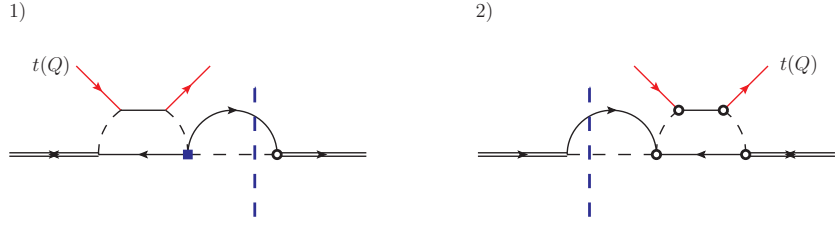}
\caption{We show diagram $a)$ of figure~\ref{fig:matchtop} and its complex conjugate 
with cuts on the lepton and the Higgs boson lines.}
\label{fig:singletop} 
\end{figure}

We denote diagrams contributing to \eqref{C0} and \eqref{C1} with $\mathcal{D}_t$ and $\mathcal{D}_Q$ respectively. 
We, first, consider diagram $a)$ of figure~\ref{fig:matchtop}. 
In this case, we can perform only one cut through the lepton line as shown in figure~\ref{fig:singletop}. 
The results read
\begin{eqnarray}
&& {\rm{Im}}\, (-i \mathcal{D}^{\ell}_{t,1,\hbox{\tiny fig.\ref{fig:singletop}}})+{\rm{Im}}\, (-i \mathcal{D}^{\ell}_{t,2,\hbox{\tiny fig.\ref{fig:singletop}}}) =
\nonumber\\
&& \hspace{3.7 cm} -\frac{|\lambda_t|^2}{M_iM_1^2} \frac{{\rm{Im}}\left[ (F_{1}^{*}F_{i})(F^*_{f1}F_{fi})\right] }{(16 \pi)^2}   
\delta^{\mu \nu} \left( P_{L} \gamma^{0} \right)^{\sigma \lambda} q_{0} + \dots \, ,  
\nonumber \\
\phantom{x}
\label{top1}\\
&& {\rm{Im}}\, (-i \mathcal{D}^{\ell}_{Q,1,\hbox{\tiny fig.\ref{fig:singletop}}})+{\rm{Im}}\, (-i \mathcal{D}^{\ell}_{Q,2,\hbox{\tiny fig.\ref{fig:singletop}}}) =
\nonumber\\
&& \hspace{3.7 cm}  -\frac{|\lambda_t|^2}{M_iM_1^2} \frac{{\rm{Im}} \left[ (F_{1}^{*}F_{i})(F^*_{f1}F_{fi}) \right] }{2(16 \pi)^2}   
\delta^{\mu \nu} \delta_{mn} \left( P_{R} \gamma^{0} \right)^{\sigma \lambda} q_{0} + \dots \, , 
\nonumber \\
\phantom{x}
\label{top2}
\end{eqnarray} 
where the dots stand for terms irrelevant for the CP asymmetry, for higher-order terms in the neutrino mass expansion 
and for terms that depend on the spin coupling of the heavy Majorana neutrino with the medium.
These last ones do not contribute if the medium is isotropic, as it is assumed in this work.

\begin{figure}[ht] 
\centering
\includegraphics[scale=0.5]{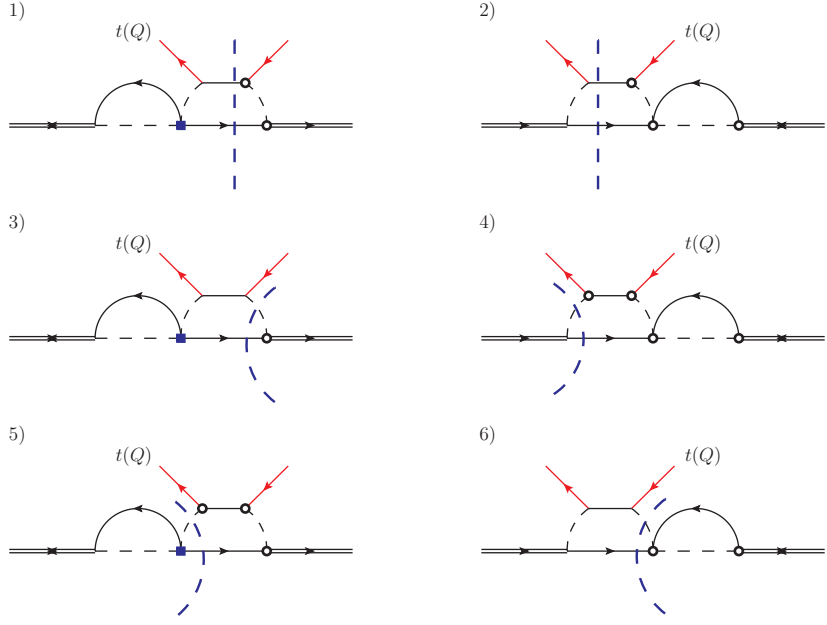}
\caption{We show diagram $b)$ of figure~\ref{fig:matchtop} and its complex conjugate 
with cuts on the lepton line and a heavy-quark doublet (top-quark) or Higgs boson line.}
\label{fig:tripletop} 
\end{figure}

We, then, consider diagram $b)$ of figure~\ref{fig:matchtop}. 
In this case the lepton line can be cut in the three different ways shown in figure~\ref{fig:tripletop}. 
Although contributions coming from single cuts may be infrared divergent, the sum of all cuts is finite.
The results read
\begin{eqnarray}
&& \hspace{-1cm}
\sum_{n=1}^{6} {\rm{Im}}\, (-i \mathcal{D}^{\ell}_{t,n,\hbox{\tiny fig.\ref{fig:tripletop}}}) = 
-\frac{3}{2}\frac{|\lambda_t|^2}{M_iM_1^2} \frac{{\rm{Im}} \left[ (F_{1}^{*}F_{i})(F^*_{f1}F_{fi}) \right] }{(16 \pi)^2}   
\delta^{\mu \nu}  \left( P_{L} \gamma^{0} \right)^{\sigma \lambda} q_{0} + \dots \, , 
\label{top1_3cut}
\end{eqnarray}
\begin{eqnarray}
&& \hspace{-1cm}
\sum_{n=1}^{6} {\rm{Im}}\, (-i \mathcal{D}^{\ell}_{Q,n,\hbox{\tiny fig.\ref{fig:tripletop}}}) = 
-\frac{3}{4}\frac{|\lambda_t|^2}{M_iM_1^2} \frac{{\rm{Im}} \left[ (F_{1}^{*}F_{i})(F^*_{f1}F_{fi}) \right] }{(16 \pi)^2}   
\delta^{\mu \nu}  \delta_{mn} \left( P_{R} \gamma^{0} \right)^{\sigma \lambda} q_{0} + \dots \, , 
\label{top2_3cut}
\end{eqnarray}
where the dots stand for terms irrelevant for the CP asymmetry and powers of $q^0/M_1$ not contributing to the matching 
of the dimension-seven operators \eqref{Ope_t} and \eqref{Ope_Q}. 

In the EFT$_2$ the matrix element \eqref{C0} reads (assuming an isotropic medium) 
\begin{equation}
\frac{{\rm{Im}} \, c^{\ell}_{3,f}}{M_1^3} \delta^{\mu \nu}  \left( P_{L} \gamma^{0} \right)^{\sigma \lambda} q_{0} \, ,
\label{eft2top}
\end{equation}
and the matrix element \eqref{C1}
\begin{equation}
\frac{{\rm{Im}} \, c^{\ell}_{4,f}}{M_1^3} \delta^{\mu \nu}  \delta_{mn} \left( P_{R} \gamma^{0} \right)^{\sigma \lambda} q_{0} \, .
\label{eft2doublet}
\end{equation}
Comparing the sum of \eqref{top1} and \eqref{top1_3cut} with \eqref{eft2top}, 
and the sum of \eqref{top2} and \eqref{top2_3cut} with \eqref{eft2doublet} 
we obtain \eqref{match_ct} and \eqref{match_cQ} respectively.
The result for anti-leptonic decays may be obtained from the substitution $F_1 \leftrightarrow F_i$ in the above expressions, 
which leads to an overall sign change in the expression of the corresponding Wilson coefficients.

We finally match the operators~\eqref{Ope_L}.
This requires computing in the EFT$_1$ the following matrix element 
\begin{equation}
-i \left.\int d^{4}x\,e^{i p \cdot x} \int d^{4}y \int d^{4}z\,e^{i q \cdot (y-z)}\, 
\langle \Omega | T(\psi^{\mu}(x) \bar{L}^{\lambda}_{h,m}(z) L^{\sigma}_{h',n}(y) \bar{\psi}^{\nu }(0))  | \Omega \rangle 
\right|_{p^\alpha =(M_1 + i\epsilon,\vec{0}\,)}, 
\label{C2}
\end{equation}
where $h$ and $h'$ are flavor indices, $\mu$, $\nu$, $\sigma$ and $\lambda$ are Lorentz indices, and $m$ and $n$ SU(2) indices. 
The matrix element \eqref{C2} describes a $2 \rightarrow 2$ scattering between a heavy Majorana neutrino at rest 
and a lepton doublet carrying momentum $q^\mu$. 

\begin{figure}[ht]
\centering
\includegraphics[scale=0.55]{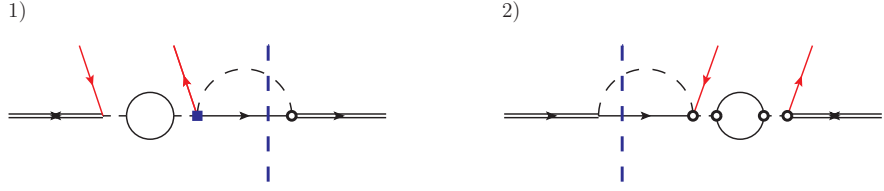}
\caption{Diagrams proportional to $|\lambda_t|^2$ contributing to the matching of the first operator~\eqref{Ope_L} 
with cuts on the lepton and the Higgs boson lines. Each diagram is the complex conjugate of the other.
The closed continuous loop is a top-quark-heavy-quark-doublet loop. 
Lepton doublets as external legs are shown as solid red lines.}
\label{fig:lepto_cut} 
\end{figure}

We consider only the diagrams proportional to  $|\lambda_t|^2$.
Differently from the diagrams discussed so far, we have to treat separately the diagrams that admit a cut on a lepton line 
from those that allow for a cut on an anti-lepton line: 
leptonic cuts contribute to $\left( \bar{N} P_{R} \, i v\cdot D L^{c}_{h'} \right) \left( \bar{L}^{c}_{h} P_{L} N \right)$ only, 
whereas cuts on anti-leptons contribute to $\left(\bar{N}P_{L} \, iv\cdot D L_{h}\right)$  $\times \left(\bar{L}_{h'} P_{R} N \right)$ only. 
We start from the diagrams in figure~\ref{fig:lepto_cut}, where we can select a lepton in the final state.\footnote{ 
A diagram similar to diagram $c)$ of figure~\ref{fig:noCP}, but with external leptons instead of Higgs bosons, 
does not contribute for the same reason as that diagram does not contribute.}
The result reads
\begin{eqnarray}
&&{\rm{Im}}\, (-i \mathcal{D}^{\ell}_{1,\hbox{\tiny fig.\ref{fig:lepto_cut}}}) +{\rm{Im}}\, (-i \mathcal{D}^{\ell}_{2,\hbox{\tiny fig.\ref{fig:lepto_cut}}})  = 
\nonumber \\
&&-\frac{9|\lambda_t|^2}{(16 \pi)^2} \, {\rm{Im}} \left[ (F_{f1}^*F_{fi}) (F^*_{h1}F_{h'i}) - (F_{f1}F^*_{fi}) (F_{h'1}F^*_{hi}) \right]  
\frac{q_0}{M_1^2 M_i} (C\,P_{R})^{\mu \sigma}(P_{L}\,C)^{\lambda \nu} \, \delta_{mn} \,.   
\nonumber \\
\phantom{x}
\label{lepto_1}
\end{eqnarray}
The combination of Yukawa couplings ${\rm{Im}}[(F_1^*F_i)(F_{f1}^*F_{fi})]$ is recovered in the CP asymmetry after computing 
the lepton tadpole in \eqref{condensate2}.
In the EFT$_2$ the leptonic contribution to the matrix element \eqref{C2} reads
\begin{equation}
\frac{{\rm{Im}}(c_{1c,f}^{hh',\ell})}{M_1^3} \, q_0 \, (C\,P_{R})^{\mu \sigma}(P_{L}\,C)^{\lambda \nu} \, \delta_{mn} \, .
\label{leptoEFT2}
\end{equation}
Comparing \eqref{lepto_1} with \eqref{leptoEFT2} we obtain the first coefficient in \eqref{match_cL}.  

\begin{figure}[ht]
\centering
\includegraphics[scale=0.55]{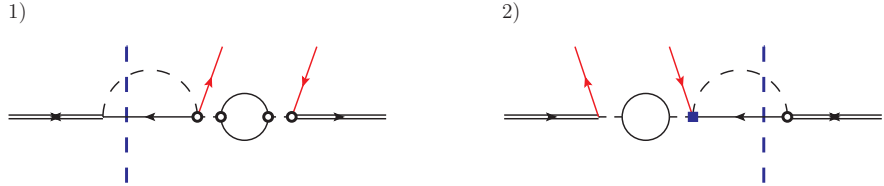}
\caption{Diagrams proportional to $|\lambda_t|^2$ contributing to the matching of the second operator~\eqref{Ope_L} 
with cuts on the anti-lepton and the Higgs boson lines. The rest is as in figure~\ref{fig:lepto_cut}.}
\label{fig:lepto_cut2} 
\end{figure}

Contributions from decays into anti-leptons come from the diagrams in figure~\ref{fig:lepto_cut2} by cutting on anti-lepton lines.
They go into the Wilson coefficient $c_{1,f}^{hh',\bar\ell}$ of the second operator~\eqref{Ope_L}.
The result reads 
\begin{eqnarray}
&&{\rm{Im}}\, (-i \mathcal{D}^{\bar\ell}_{1,\hbox{\tiny fig.\ref{fig:lepto_cut2}}}) +{\rm{Im}}\, (-i \mathcal{D}^{\bar\ell}_{2,\hbox{\tiny fig.\ref{fig:lepto_cut2}}}) =  
\nonumber\\
&&-\frac{9|\lambda_t|^2}{(16 \pi)^2} \, {\rm{Im}} \left[ (F_{f1}F_{fi}^*) (F_{h'1}F^*_{hi}) - (F_{f1}^*F_{fi}) (F^*_{h1}F_{h'i}) \right]  
\frac{q_0}{M_1^2 M_i} (P_{L})^{\mu \lambda}(P_{R})^{\sigma \nu} \, \delta_{mn} \, .
\nonumber\\
\phantom{x}
\label{lepto_2} 
\end{eqnarray}
In the EFT$_2$ the anti-leptonic contribution to the matrix element \eqref{C2} reads
\begin{equation}
\frac{{\rm{Im}}(c_{1,f}^{hh',\bar\ell})}{M_1^3} \, q_0 \, (P_{L})^{\mu \lambda}(P_{R})^{\sigma \nu} \, \delta_{mn}\, .
\label{leptoEFT2_2}
\end{equation}
Comparing \eqref{lepto_2} with \eqref{leptoEFT2_2} we obtain the second coefficient in \eqref{match_cL}.


\newpage


\begin{thebibliography}{99}

\bibitem{Yoshimura:1978ex}
  M.~Yoshimura,
  Phys.\ Rev.\ Lett.\  {\bf 41} (1978) 281
   Erratum: [Phys.\ Rev.\ Lett.\  {\bf 42} (1979) 746].

\bibitem{Toussaint:1978br}
  D.~Toussaint, S.~B.~Treiman, F.~Wilczek and A.~Zee,
  Phys.\ Rev.\ D {\bf 19} (1979) 1036.
 
\bibitem{Weinberg:1979bt}
  S.~Weinberg,
  Phys.\ Rev.\ Lett.\  {\bf 42} (1979) 850.
  
\bibitem{Dimopoulos:1978kv}
  S.~Dimopoulos and L.~Susskind,
  Phys.\ Rev.\ D {\bf 18} (1978) 4500.
  
\bibitem{Sakharov:1967dj}
  A.~D.~Sakharov,
  Pisma Zh.\ Eksp.\ Teor.\ Fiz.\  {\bf 5} (1967) 32
   [JETP Lett.\  {\bf 5} (1967) 24]
   [Sov.\ Phys.\ Usp.\  {\bf 34} (1991) 392]
   [Usp.\ Fiz.\ Nauk {\bf 161} (1991) 61].
  
\bibitem{Giudice:2000ex}
  G.~F.~Giudice, E.~W.~Kolb and A.~Riotto,
  Phys.\ Rev.\ D {\bf 64} (2001) 023508
  [hep-ph/0005123].
  
\bibitem{Blanchet:2012bk}
  S.~Blanchet and P.~Di Bari,
  New J.\ Phys.\  {\bf 14} (2012) 125012
  [arXiv:1211.0512 [hep-ph]].
  
\bibitem{Fukugita:1986hr}
  M.~Fukugita and T.~Yanagida,
  Phys.\ Lett.\ B {\bf 174} (1986) 45.
  
\bibitem{Kuzmin:1985mm}
  V.~A.~Kuzmin, V.~A.~Rubakov and M.~E.~Shaposhnikov,
  Phys.\ Lett.\ B {\bf 155} (1985) 36.
  
\bibitem{Minkowski:1977sc}
  P.~Minkowski,
  Phys.\ Lett.\ B {\bf 67} (1977) 421.
  
\bibitem{GellMann:1980vs}
  M.~Gell-Mann, P.~Ramond and R.~Slansky,
  Conf.\ Proc.\ C {\bf 790927} (1979) 315
  [arXiv:1306.4669 [hep-th]].

\bibitem{Mohapatra:1979ia}
  R.~N.~Mohapatra and G.~Senjanovic,
  Phys.\ Rev.\ Lett.\  {\bf 44} (1980) 912.
  
\bibitem{Fukuda:1998mi}
  Y.~Fukuda {\it et al.} [Super-Kamiokande Collaboration],
  Phys.\ Rev.\ Lett.\  {\bf 81} (1998) 1562
  [hep-ex/9807003].

\bibitem{Buchmuller:2004nz}
 W.~Buchm\"uller, P.~Di Bari and M.~Pl\"umacher,
 Annals Phys.\  {\bf 315} (2005) 305
 [hep-ph/0401240].
  
\bibitem{Fong:2013wr}
  C.~S.~Fong, E.~Nardi and A.~Riotto,
  Adv.\ High Energy Phys.\  {\bf 2012} (2012) 158303
  [arXiv:1301.3062 [hep-ph]].
  
\bibitem{Buchmuller:2005eh}
  W.~Buchm\"uller, R.~D.~Peccei and T.~Yanagida,
  Ann.\ Rev.\ Nucl.\ Part.\ Sci.\  {\bf 55} (2005) 311
  [hep-ph/0502169].

\bibitem{Drewes:2013gca}
  M.~Drewes,
  Int.\ J.\ Mod.\ Phys.\ E {\bf 22} (2013) 1330019
  [arXiv:1303.6912 [hep-ph]].
  
\bibitem{Biondini:2013xua}
  S.~Biondini, N.~Brambilla, M.~A.~Escobedo and A.~Vairo,
  JHEP {\bf 1312} (2013) 028
  [arXiv:1307.7680].
  
\bibitem{Biondini:2015gyw}
  S.~Biondini, N.~Brambilla, M.~A.~Escobedo and A.~Vairo,
  JHEP {\bf 1603} (2016) 191
  [arXiv:1511.02803 [hep-ph]].

\bibitem{Covi:1997dr}
  L.~Covi, N.~Rius, E.~Roulet and F.~Vissani,
  Phys.\ Rev.\ D {\bf 57} (1998) 93
  [hep-ph/9704366].

\bibitem{Giudice:2003jh}
  G.~F.~Giudice, A.~Notari, M.~Raidal, A.~Riotto and A.~Strumia,
  Nucl.\ Phys.\ B {\bf 685} (2004) 89
  [hep-ph/0310123].

\bibitem{Garny:2010nj}
  M.~Garny, A.~Hohenegger and A.~Kartavtsev,
  Phys.\ Rev.\ D {\bf 81} (2010) 085028
  [arXiv:1002.0331 [hep-ph]].

\bibitem{Anisimov:2010dk}
  A.~Anisimov, W.~Buchm\"uller, M.~Drewes and S.~Mendizabal,
  Annals Phys.\  {\bf 326} (2011) 1998
   [Annals Phys.\  {\bf 338} (2011) 376]
  [arXiv:1012.5821 [hep-ph]].

\bibitem{Kiessig:2011fw}
  C.~Kiessig and M.~Pl\"umacher,
  JCAP {\bf 1207} (2012) 014
  [arXiv:1111.1231 [hep-ph]].
  
\bibitem{Nardi:2005hs}
  E.~Nardi, Y.~Nir, J.~Racker and E.~Roulet,
  JHEP {\bf 0601} (2006) 068
  [hep-ph/0512052].  
  
\bibitem{Nardi:2006fx}
  E.~Nardi, Y.~Nir, E.~Roulet and J.~Racker,
  JHEP {\bf 0601} (2006) 164
  [hep-ph/0601084].  
  
\bibitem{Davidson:2002qv}
  S.~Davidson and A.~Ibarra,
  Phys.\ Lett.\ B {\bf 535} (2002) 25
  [hep-ph/0202239].
  
\bibitem{Buchmuller:2002rq}
  W.~Buchm\"uller, P.~Di Bari and M.~Pl\"umacher,
  Nucl.\ Phys.\ B {\bf 643} (2002) 367
   Erratum: [Nucl.\ Phys.\ B {\bf 793} (2008) 362]
  [hep-ph/0205349].

\bibitem{Racker:2012vw}
  J.~Racker, M.~Pena and N.~Rius,
  JCAP {\bf 1207} (2012) 030
  [arXiv:1205.1948 [hep-ph]].
  
\bibitem{Kawasaki:2004qu}
  M.~Kawasaki, K.~Kohri and T.~Moroi,
  Phys.\ Rev.\ D {\bf 71} (2005) 083502
  [astro-ph/0408426].
  
\bibitem{Buchmuller:2000nd}
  W.~Buchm\"uller and S.~Fredenhagen,
  Phys.\ Lett.\ B {\bf 483} (2000) 217
  [hep-ph/0004145].
  
\bibitem{Laine:2013lka}
  M.~Laine,
  JHEP {\bf 1308} (2013) 138
  [arXiv:1307.4909 [hep-ph]].
  
\bibitem{Salvio:2011sf}
  A.~Salvio, P.~Lodone and A.~Strumia,
  JHEP {\bf 1108} (2011) 116
  [arXiv:1106.2814 [hep-ph]].

\bibitem{Laine:2011pq}
  M.~Laine and Y.~Schr\"oder,
  JHEP {\bf 1202} (2012) 068
  [arXiv:1112.1205 [hep-ph]].
  
\bibitem{Bodeker:2013qaa}
  D.~B\"odeker and M.~W\"ormann,
  JCAP {\bf 1402} (2014) 016
  [arXiv:1311.2593 [hep-ph]].
  
\bibitem{Liu:1993tg}
  J.~Liu and G.~Segre,
  Phys.\ Rev.\ D {\bf 48} (1993) 4609
  [hep-ph/9304241].
  
\bibitem{Covi:1996wh}
  L.~Covi, E.~Roulet and F.~Vissani,
  Phys.\ Lett.\ B {\bf 384} (1996) 169
  [hep-ph/9605319].

\bibitem{Boyarsky:2009ix}
  A.~Boyarsky, O.~Ruchayskiy and M.~Shaposhnikov,
  Ann.\ Rev.\ Nucl.\ Part.\ Sci.\  {\bf 59} (2009) 191
  [arXiv:0901.0011 [hep-ph]].

\bibitem{Buttazzo:2013uya}
  D.~Buttazzo, G.~Degrassi, P.~P.~Giardino, G.~F.~Giudice, F.~Sala, A.~Salvio and A.~Strumia,
  JHEP {\bf 1312} (2013) 089
  [arXiv:1307.3536 [hep-ph]].
   
\bibitem{Rose15}
 L.~Delle Rose, private communication.  

\bibitem{Brambilla:2003nt}
  N.~Brambilla, D.~Gromes and A.~Vairo,
  Phys.\ Lett.\ B {\bf 576} (2003) 314
  [hep-ph/0306107].
  
\bibitem{Luty:1992un}
  M.~A.~Luty,
  Phys.\ Rev.\ D {\bf 45} (1992) 455.
  
\bibitem{Plumacher:1996kc}
  M.~Pl\"umacher,
  Z.\ Phys.\ C {\bf 74} (1997) 549
  [hep-ph/9604229].

\bibitem{Braaten:1989mz}
  E.~Braaten and R.~D.~Pisarski,
  Nucl.\ Phys.\ B {\bf 337} (1990) 569.
  
\bibitem{Shtabovenko:2016sxi}
  V.~Shtabovenko, R.~Mertig and F.~Orellana,
  arXiv:1601.01167 [hep-ph].
  
\bibitem{Vlad2}
  V.~Shtabovenko, FeynHelpers: Connecting FeynCalc to FIRE and Package-X, TUM-EFT 75/15, in preparation.  
  
\bibitem{Cutkosky:1960sp}
  R.~E.~Cutkosky,
  J.\ Math.\ Phys.\  {\bf 1} (1960) 429.

\bibitem{Remiddi:1981hn}
  E.~Remiddi,
  Helv.\ Phys.\ Acta {\bf 54} (1982) 364.
  
\bibitem{Bellac}
  M.~Le Bellac,
  Quantum and Statistical Field Theory, (Oxford University Press, 1991).
  
\bibitem{Denner:2014zga}
  A.~Denner and J.~N.~Lang,
  Eur.\ Phys.\ J.\ C {\bf 75} (2015) 8,  377
  [arXiv:1406.6280 [hep-ph]].
    
\end{thebibliography}
\end{document}